\documentclass[10pt]{article}
\usepackage{arxiv}

\usepackage[utf8x]{inputenc}
\usepackage{multirow}
\usepackage{multicol}
\usepackage{comment}
\usepackage{todonotes}
\usepackage{url}
\usepackage{csquotes}       
\usepackage{caption}        
\usepackage{subcaption}     
\usepackage{pgfplots,siunitx}	
\usetikzlibrary{patterns.meta,pgfplots.groupplots, backgrounds, shapes, arrows.meta}
\usepgfplotslibrary{fillbetween}
\pgfplotsset{compat=1.4, 
legend image code/.code={
\draw[mark repeat=2,mark phase=2]
plot coordinates {
(0cm,0cm)
(0.15cm,0cm)        
(0.3cm,0cm)         
};%
}
}
\usepackage{tikzscale}	
\usepackage{tikz-layers}    

\usepackage{pifont}
\newcommand{\cmark}{\ding{51}}%
\newcommand{\xmark}{\ding{55}}%
\usepackage{hwemoji}  

\pgfdeclarelayer{background}
\pgfdeclarelayer{foreground}
\pgfsetlayers{background,main,foreground}  

\usepackage[natbib,style=ieee,defernumbers=true]{biblatex}
\DeclarePrintbibliographyDefaults{title=\refname}
\AtEveryBibitem{%
\clearfield{doi}%
\clearfield{isbn}%
\clearfield{issn}%
\clearfield{urldate}%
\clearfield{urlyear}%
\clearfield{urlmonth}%
\clearfield{pages}%
}

\newcommand*\circled[2]{\tikz[baseline=(char.base)]{
\node[circle,fill=white,draw,scale=#2,inner sep=1.0pt,text=black] (char) {#1};}}

\newif\ifremark
\long\def\remark#1{
\ifremark%
        \begingroup%
        \dimen0=\columnwidth
        \advance\dimen0 by -1in%
        \setbox0=\hbox{\parbox[b]{\dimen0}{\protect\em #1}}
        \dimen1=\ht0\advance\dimen1 by 2pt%
        \dimen2=\dp0\advance\dimen2 by 2pt%
        \vskip 0.25pt%
        \hbox to \columnwidth{%
                \vrule height\dimen1 width 3pt depth\dimen2%
                \hss\copy0\hss%
                \vrule height\dimen1 width 3pt depth\dimen2%
        }%
        \endgroup%
\fi}
\remarktrue

\newcommand{\ignore}[1]{}

\newcommand{\schemename}[1]{ST-SP}
\newcommand{\platformname}[1]{\mbox{EStacker}}

\title{EStacker: Explaining Battery-Less IoT System Performance with Energy Stacks}

\author{
    Lukas Liedtke \\
    Norwegian University of Science and Technology\\
    Trondheim, Norway \\
    \texttt{lukas.liedtke@ntnu.no} \\
    \And
    Per Gunnar Kjeldsberg\\
    Norwegian University of Science and Technology\\
    Trondheim, Norway \\
    \texttt{pgk@ntnu.no} \\
    \And
    Frank Alexander Kraemer\\
    Norwegian University of Science and Technology\\
    Trondheim, Norway \\
    \texttt{kraemer@ntnu.no} \\
    \And
    Magnus Jahre\\
    Norwegian University of Science and Technology\\
    Trondheim, Norway \\
    \texttt{magnus.jahre@ntnu.no} \\
}

\begin{document}

\maketitle

\begin{abstract}
The number of Internet of Things (IoT) devices is increasing exponentially, and it is environmentally and economically unsustainable to power all these devices with batteries. The key alternative is energy harvesting, but battery-less IoT systems require extensive evaluation to demonstrate that they are sufficiently performant across the full range of expected operating conditions.
IoT developers thus need an evaluation platform that (i) ensures that each evaluated application and configuration is exposed to exactly the same energy environment and events, and (ii) provides a detailed account of what the application spends the harvested energy on.
We therefore developed the EStacker evaluation platform which (i) provides fair and repeatable evaluation, and (ii) generates energy stacks. Energy stacks break down the total energy consumption of an application across hardware components and application activities, thereby explaining what the application specifically uses energy on.
We augment EStacker with the ST-SP optimization which, in our experiments, reduces evaluation time by 6.3$\times$ on average while retaining the temporal behavior of the battery-less IoT system (average throughput error of 7.7\%) by proportionally scaling time and power.
We demonstrate the utility of EStacker through two case studies. In the first case study, we use energy stack profiles to identify a performance problem that, once addressed, improves performance by 3.3$\times$. The second case study focuses on ST-SP, and we use it to explore the design space required to dimension the harvester and energy storage sizes of a smart parking application in roughly one week (7.7 days). Without ST-SP, sweeping this design space would have taken well over one month (41.7 days).
\end{abstract}


\section{Introduction}
\label{sec:intro}

The number of Internet of Things (IoT) devices has increased significantly over the last decade and will continue to grow with almost 40 billion IoT connections predicted for 2033~\cite{mattarnott:2024:global}. IoT applications are also becoming increasingly diverse with recent examples ranging from environment and asset monitoring~\cite{zaghari_high_temperature_2020,giordano:2022:design} and medical implants~\cite{zhang:2022:battery} to computational pipelines in space~\cite{denby_orbital_2020,denby:2022:tartan}. Most IoT devices are powered by chemical batteries which is problematic because it can be costly, and sometimes even impossible, to replace them (e.g., in space or within the human body). Moreover, used batteries can have a significant environmental impact, especially when considering that the number of batteries disposed of daily is expected to reach hundreds of millions~\cite{hayes_research_2021} and that the footprint of the battery is a dominating factor in the overall carbon footprint of IoT devices~\cite{prakash_is_2023}. It is hence attractive to design battery-less IoT systems that use Energy Harvesting~(EH) to collect energy from their environment, for example using solar~\cite{denby:2022:tartan}, motion~\cite{gorlatova:2015:movers}, temperature differences~\cite{afanasov:2020:batteryless}, or radio waves~\cite{sample:2008:design} as energy sources.

To be a viable alternative to battery-powered systems, developers must be able to demonstrate that their IoT application meets performance requirements and constraints on the battery-less device they target. This is challenging because the behavior of battery-less systems, and ultimately their performance, can vary significantly when exposed to different energy harvesting conditions~\cite{hester:2017:realistic}. Moreover, IoT applications are often (a combination of) {\it periodic} --- i.e., they collect, process, and communicate sensor data at fixed time intervals --- and {\it reactive} --- i.e., they communicate as a consequence of detecting changes to their environment --- and a viable evaluation approach must cater to the full range of the application spectrum spanned by the combinations of these broad approaches. Developers must expose the hardware and software configurations they want to explore to exactly the same energy harvesting conditions to ensure that comparisons are fair --- i.e., that system A performs better than system B because of its superior technical qualities and not because of differences in the amount or timing of supplied energy. For reactive applications, developers must also control the relevant aspects of the environment, such as temperature exceeding a threshold, because exposing configurations to different environmental conditions will change system behavior and performance, rendering system comparisons inconclusive.

\begin{figure}[t]
    \begin{center}    
    \hspace{0.0cm}\includegraphics[width=\columnwidth, trim=0cm 10cm 0cm 2.7cm, clip]{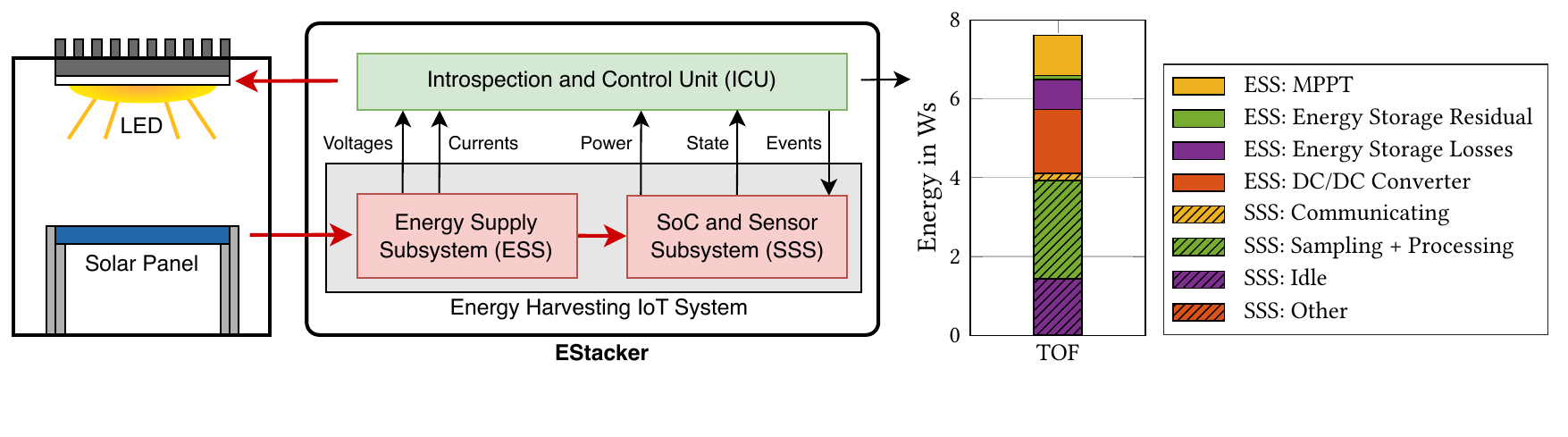}
    \end{center}
    \vspace*{-0.6cm}
    \caption{Our EStacker evaluation platform for battery-less IoT systems. {\it EStacker enables repeatable evaluation of periodic and reactive IoT applications and generates energy stacks that break down energy consumption across hardware components and application activities.}}
    \label{Fig:HighLvl_EvalSetup}
\end{figure}

IoT developers hence need an evaluation platform that enables repeatable evaluation of periodic and reactive IoT applications on battery-less devices as this ensures that end-to-end performance metrics such as execution time or sampling bandwidth can be compared fairly and reliably. While this is a necessary first step, designers also need to understand {\it why} one configuration outperforms another. In systems where energy is abundant, the fundamental performance metric is execution time~\cite{hennessy_computer_2018}, and the go-to strategy to glean insight into performance is to break down overall execution time into the key activities of the system, resulting in various forms of (clock) cycle stacks~\cite{yasin_top-down_2014, eyerman_multi-stage_2018, eyerman_performance_2006, gottschall_per-instruction_2024, eyerman_speedup_2012}.
The key benefit of cycle stacks is that they clearly explain what an application spends time on and thereby identify the subsystem for which optimization, either in software or hardware, is likely to have the greatest impact on overall performance (due to Amdahl's law). In a battery-less system, the best-performing configuration will be the one that uses the harvested energy most efficiently, and an effective evaluation platform should thus provide {\it energy stacks.} An energy stack, as exemplified in Figure~\ref{Fig:HighLvl_EvalSetup}, breaks down the total amount of harvested energy across the hardware components and application activities it is consumed by, thereby explaining what the IoT device spends energy on and highlighting opportunities for energy optimization.

The need for repeatable evaluation of battery-less systems is well known, and EHTestbed~\cite{sigrist_environment_2021}, Shepherd~\cite{geissdoerfer:2019:shepherd}, and Ekho~\cite{hester:2017:realistic} are the state-of-the-art evaluation platforms that aim to address this problem. While they support periodic and reactive applications to varying degrees,\footnote{EHTestbed, Shepherd, and Ekho all support periodic applications. EHTestbed and Shepherd support reactive applications in load emulation mode and through an external command interface, respectively, while Ekho does not support reactive applications. We refer to the platform proposed by Sigrist et al.~\cite{sigrist_environment_2021} as EHTestbed, but acknowledge that the original authors did not name their platform.} they cannot create energy stacks. The reason is that energy stacks require extensive out-of-band measurements of each hardware component's energy consumption which is impractical, or even impossible, to add to an existing platform. More specifically, it would require adding additional components and redesigning the Printed Circuit Board (PCB) to the extent that the result would be better described as a new platform. Standalone power measurement devices such as eProfiler~\cite{kazdaridis:2020:eprofiler} and RocketLogger~\cite{sigrist_measurement_2017} cannot create energy stacks for the same reason, i.e., they cannot access power supply lines that the platform does not expose.

We hence designed EStacker to fill this gap and thereby created the first evaluation platform for battery-less systems that (i) controls the energy and event environments, {\it and} (ii) generates energy stacks.  
Figure~\ref{Fig:HighLvl_EvalSetup} introduces EStacker in the context of a solar-powered battery-less IoT system. EStacker emulates the environment based on traces, e.g., irradiance traces for the solar harvester and event traces for sensor events. EStacker's {\it Introspection and Control Unit (ICU)} controls the amount of energy supplied to the {\it EH IoT System} by using the irradiance trace to drive an LED that provides energy to the solar panel. Similarly, the ICU uses an event trace to notify the {\it SoC and Sensor Subsystem (SSS)} that an event has occurred. The {\it EH IoT System} is solely powered by harvested energy whereas the ICU has its own non-intermittent power source. The ICU also measures the energy consumption of the key components of the {\it Energy Supply Subsystem (ESS)} and SSS as well as the application activities to create energy stacks. More specifically, the energy stacks as shown on the right in Figure~\ref{Fig:HighLvl_EvalSetup} report the individual energy consumption of the Maximum-Power-Point Tracker (MPPT), the residual and lost energy in the energy storage, and the DC/DC converter within the ESS. Within the SSS, the energy stacks report the energy consumed during communication, sampling and processing, and while idle; the remaining SSS components such as (re-)boots and backups are grouped in the 'SSS: Other' category. The energy stack of the {\sf TOF} benchmark clearly shows that its key optimization target is sampling and processing as this accounts for 32.9\% of its total energy consumption. (We will explain the details of how EStacker creates energy stacks in Section~\ref{sec:platform} and present and analyze a variety of energy stacks across realistic  IoT benchmarks in Sections~\ref{sec:results} and \ref{sec:case-studies}.)

While EStacker can capture energy stacks for any application, exposing such systems to a meaningful number of relevant energy and event scenarios can result in evaluation times of days, weeks, or even months. This problem is particularly severe for earth-bound outdoor solar harvesting, due to its diurnal and weather-related (short-term) fluctuations as well as its (long-term) seasonal variations. There is hence a need to reduce evaluation times while maintaining accuracy, and we propose ST-SP to this end. In contrast to EHTestbed's \textit{Scaled Time~- Unscaled Power (ST-UP)} strategy~\cite{sigrist_environment_2021}, which only scales time, ST-SP proportionally scales time and power on both the input (harvester) and output (SoC) sides, i.e., a \textit{Scaled Time~- Scaled Power} strategy. 
On the input side, we can easily control time and power by scaling  EStacker's irradiance trace, but the output side is more challenging because we have to adjust the activity level of the SoC to scale time and power. To address this challenge, we leverage that many sensor-focused IoT applications have a degree of periodicity, for example only periodically enabling sensor(s) to save energy. ST-SP exploits this characteristic to adjust the application's period such that the relation between input and output power under scaling remains the same as when the application is executed in real time. In contrast to EStacker, the ST-SP optimization hence cannot be applied to purely reactive applications.
We evaluate ST-SP and ST-UP on EStacker across a range of IoT benchmarks and find that ST-SP predicts throughput with an absolute relative error of 7.7\%, a significant improvement over ST-UP's 21.2\% error. 

We demonstrate that EStacker is useful in practice through two case studies, see Section~\ref{sec:case-studies} for details. The first uses EStacker's ability to generate an energy stack profile, i.e., energy stacks collected at regular intervals during execution, to investigate if the {\sf TOF} benchmark from Figure~\ref{Fig:HighLvl_EvalSetup} uses energy efficiently during sampling and processing. We collected this energy stack on an earlier version of EStacker, which suffered from a flaw that resulted in the equivalent series resistance of the capacitor banks causing unexpected shutdowns for {\sf TOF}, ultimately leaving the sensor powered on when it should have been powered off. Fixing this issue enabled {\sf TOF} to instead use this energy on collecting, processing, and communicating samples, thereby improving performance by 3.3$\times$. Our second case study demonstrates how EStacker's ST-SP optimization makes it practical to sweep a large design space to identify favorable solar panel and energy storage sizes for a smart parking application. ST-SP reduces the time it takes to explore this design space from well over one month (41.7 days) to roughly one week (7.7 days).

\newpage

\noindent
In summary, we make the following major contributions:
\begin{itemize}
    \item We present \platformname{}, an evaluation platform for battery-less IoT devices that supports periodic and reactive applications {\it and} generates energy stacks, thereby explaining what the application spends energy on. 
    \item We propose ST-SP which speeds up the evaluation of battery-less IoT systems by appropriately scaling (i)~the playback speed of the energy trace, (ii) the input power in the emulated energy environment, and (iii) the sampling frequency of the periodic IoT application (i.e., its average power consumption). ST-SP speeds up evaluation by 6.3$\times$ with average throughput and activity profile errors of 7.7\% and 1.4\%, respectively --- a significant improvement over the 21.2\% and 28.6\% average errors of state-of-the-art ST-UP~\cite{sigrist_environment_2021}.
    \item We provide two case studies to demonstrate that EStacker is useful in practice. The first uses energy stack profiles to identify a performance problem that, once fixed, improves the performance of the {\sf TOF} benchmark by 3.3$\times$. In the second case study, we use the ST-SP optimization to explore a design space in roughly one week (7.7 days) that would have taken over a month (41.7 days) to explore without ST-SP.
\end{itemize}

\section{Battery-Less IoT System Architecture}

\begin{figure}[t]
    \begin{center}
    \includegraphics[width=0.65\columnwidth]{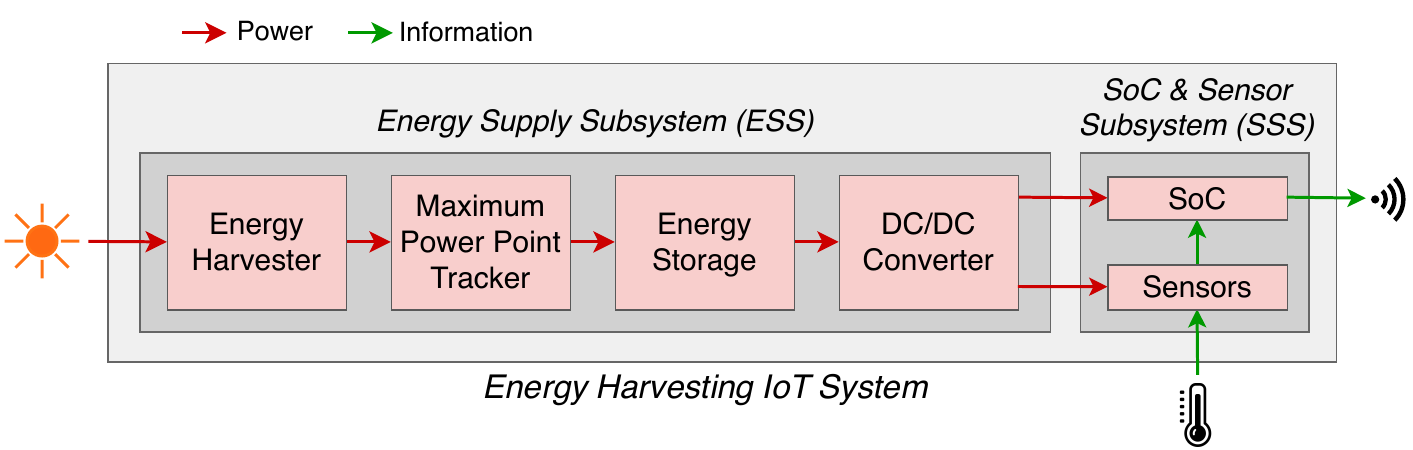}\vspace*{-2mm}
    \caption{The hardware architecture of a state-of-the-art battery-less IoT system~\cite{colin:2018:reconfigurable, ju:2018:predictive, nardello_camaroptera_2019, afanasov:2020:batteryless, ruppel:2022:architectural}.}
    \label{fig:EH_IoT_Hardware_Arch}
    \end{center}
\end{figure}

Figure~\ref{fig:EH_IoT_Hardware_Arch} illustrates a state-of-the-art battery-less IoT system~\cite{colin:2018:reconfigurable, ju:2018:predictive, nardello_camaroptera_2019, afanasov:2020:batteryless, ruppel:2022:architectural}.

The two high-level subsystems are the \textit{Energy Supply Subsystem (ESS)}, and the \textit{SoC \& Sensor Subsystem (SSS)}; these collectively form the {\it EH IoT System} in Figure~\ref{Fig:HighLvl_EvalSetup}. The SSS handles data acquisition (sampling), processing, and (wireless) communication~\cite{ghasemi_pes:_2023}, and typically consists of a low-power System-on-Chip (SoC) and various sensors (e.g., temperature, humidity, or acceleration).

The ESS consists of various components which we will shortly describe in detail. This complexity is typically necessary because (i) the available ambient energy has to be converted from its original form into electrical energy, and (ii) the available input power typically fluctuates over time. Therefore, the ESS needs to be carefully designed to match the performance requirements of the foreseen application. This, however, poses a challenge for developers because creating an efficient system requires considering the amount of energy available, how the energy supply fluctuates over time, and the expected power consumption profile of the application. Further difficulties arise from the interactions between the various components of the ESS, where design parameters, such as the voltage thresholds that determine when to start and stop operating, can significantly impact ESS efficiency. This ultimately impacts performance as the ESS must supply sufficient energy for the IoT application to satisfy its performance constraints.

The first component in the ESS is the {\it Energy Harvester}, which converts the ambient energy at the deployment location into electrical energy. In the simplest case, the energy harvester powers the SSS directly, which minimizes both circuit board size and component cost. The difficulty is that when selecting the specific system components, great care must be taken to ensure that the harvester's output voltage does not exceed the maximum operating voltage of any of the connected components. In addition, with a directly connected harvester, the system can only operate when (i) the harvester supplies more power than the downstream components consume, and (ii) its output voltage is within the operating voltage ranges of the downstream components. 
Such systems hence exhibit hard intermittency as they must be carefully designed to maintain forward progress in application execution even when frequently shutting down. While this strategy can be necessary for energy harvesters that provide limited energy, a soft intermittency strategy, in which the system attempts to avoid fully shutting down, is typically more efficient with more powerful energy harvesters~\cite{wymore_tale_2023}. An added benefit of focusing on higher-energy harvesters is that a broader range of sensors and SoCs can be supported, and we therefore primarily focus on soft intermittent systems in this paper.

The ESS should extract as much energy from the harvester as possible, but this is challenging because the output power of the harvester depends on both ambient power density (e.g., solar irradiance) and the harvester's output voltage. The relationship between output current and voltage can be described by IV-curves~\cite{hester:2017:realistic}, and, by multiplying current and output voltage and plotting against output voltage, we get a power-voltage curve. For (non-shaded) solar cells, the power-voltage curve shows a single maximum, which marks the maximum power point. Connecting the harvester directly to the energy storage therefore yields suboptimal power output most of the time because it passes through all points on the power-voltage curve during charging. For this reason, state-of-the-art platforms typically employ {\it Maximum Power Point Trackers (MPPTs)} between the energy harvester and the energy storage. These integrated circuits attempt to adjust the operating point of the harvester to maximize its power output. 

State-of-the-art ESS implementations include {\it Energy Storage}, for example in the form of \mbox{(super-)capacitors}, to save energy that the SSS does not immediately consume. Since many battery-less systems operate in environments with strongly fluctuating ambient energy, the common case is that the consumption of the SSS does not match the available power. The dimensioning of the energy storage depends on the performance requirements and the physical constraints of the IoT system. More specifically, the energy storage capacity has a significant impact on the responsiveness of the application~\cite{colin:2018:reconfigurable} and its ability to maintain operation over longer periods of energy scarcity. 

The ESS must strive to keep its output voltages within the specified operating voltage ranges of the components of the SSS. If the ESS output voltage falls below the minimum operating voltage $V_\text{min}$ of an SSS component, this component will shut down. If the SSS is connected directly to a \mbox{(super-)capacitor} that is used as energy storage, the \mbox{(super-)capacitor} will still contain $E_\text{res}$ energy when its output voltage is $V_\text{min}$; $E_\text{res}$ equals $1/2\times CV^2_\text{min}$ for capacitance $C$. The residual energy $E_\text{res}$ can hence be significant when capacitance is high. This is inefficient as $E_\text{res}$ is harvested energy that the SSS cannot make use of.

Moreover, the ESS must ensure that the voltage it provides to the components of the SSS never exceeds their maximum operating voltages as this may harm the components. For these reasons, state-of-the-art SSS implementations use {\it DC/DC converter(s)} after the energy storage to (i) guarantee that the output voltage(s) of the ESS is within the supported ranges of downstream components, (ii) allow the energy storage to be charged to a higher voltage than the downstream components support, and (iii) allow the SSS to operate for energy storage voltages below $V_\text{min}$.

\section{EStacker}
\label{sec:platform}

\begin{figure}[t]
    \begin{subfigure}{0.48\columnwidth}
        \includegraphics[width=\columnwidth, trim=0cm 0.3cm 0cm 0cm, clip]{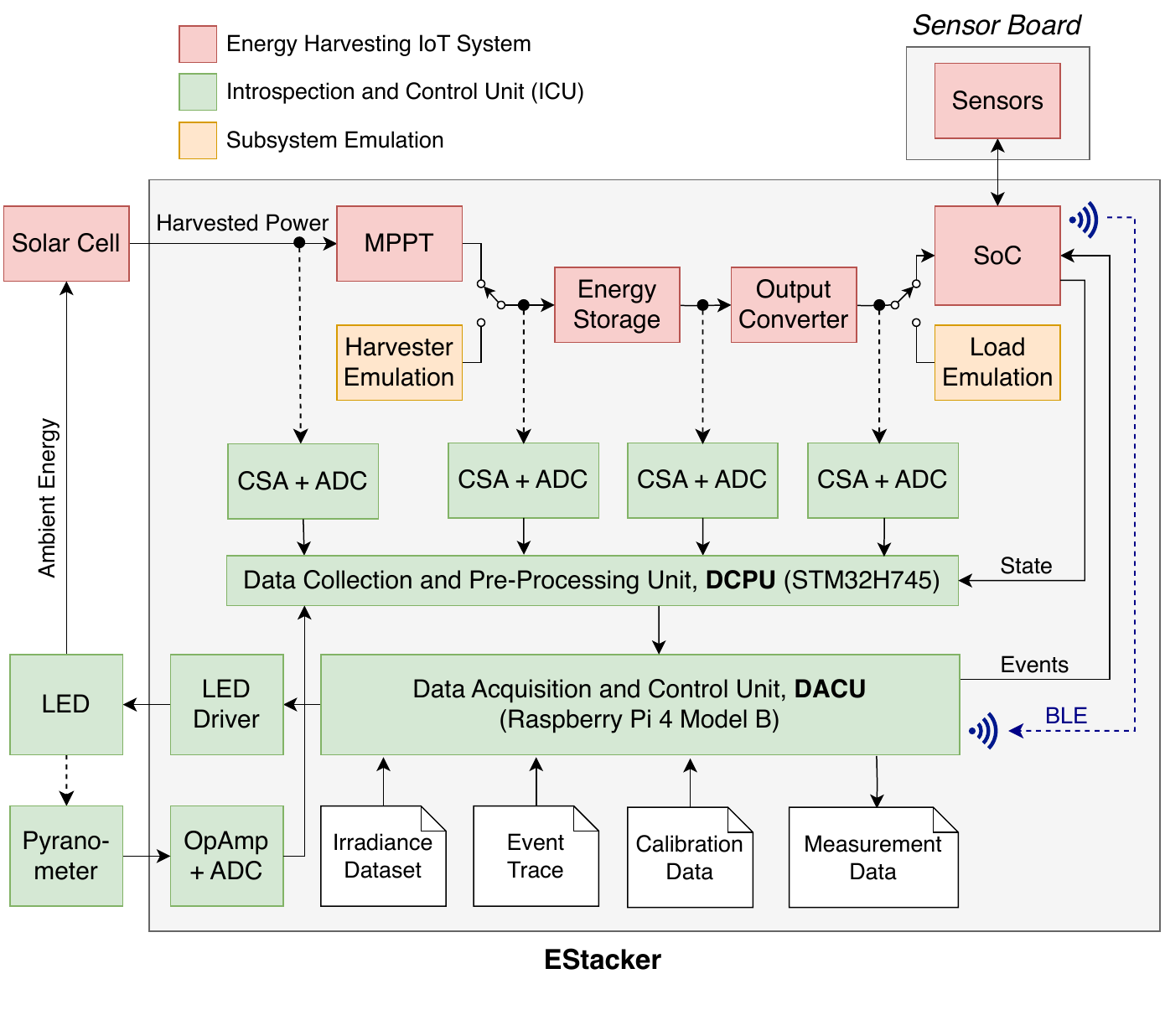}
        \caption{EStacker system architecture.}
        \label{fig:Platform_Block_Diagram}
    \end{subfigure}
    \hfill
    \begin{subfigure}{0.5\columnwidth}
        \includegraphics[width=\columnwidth]{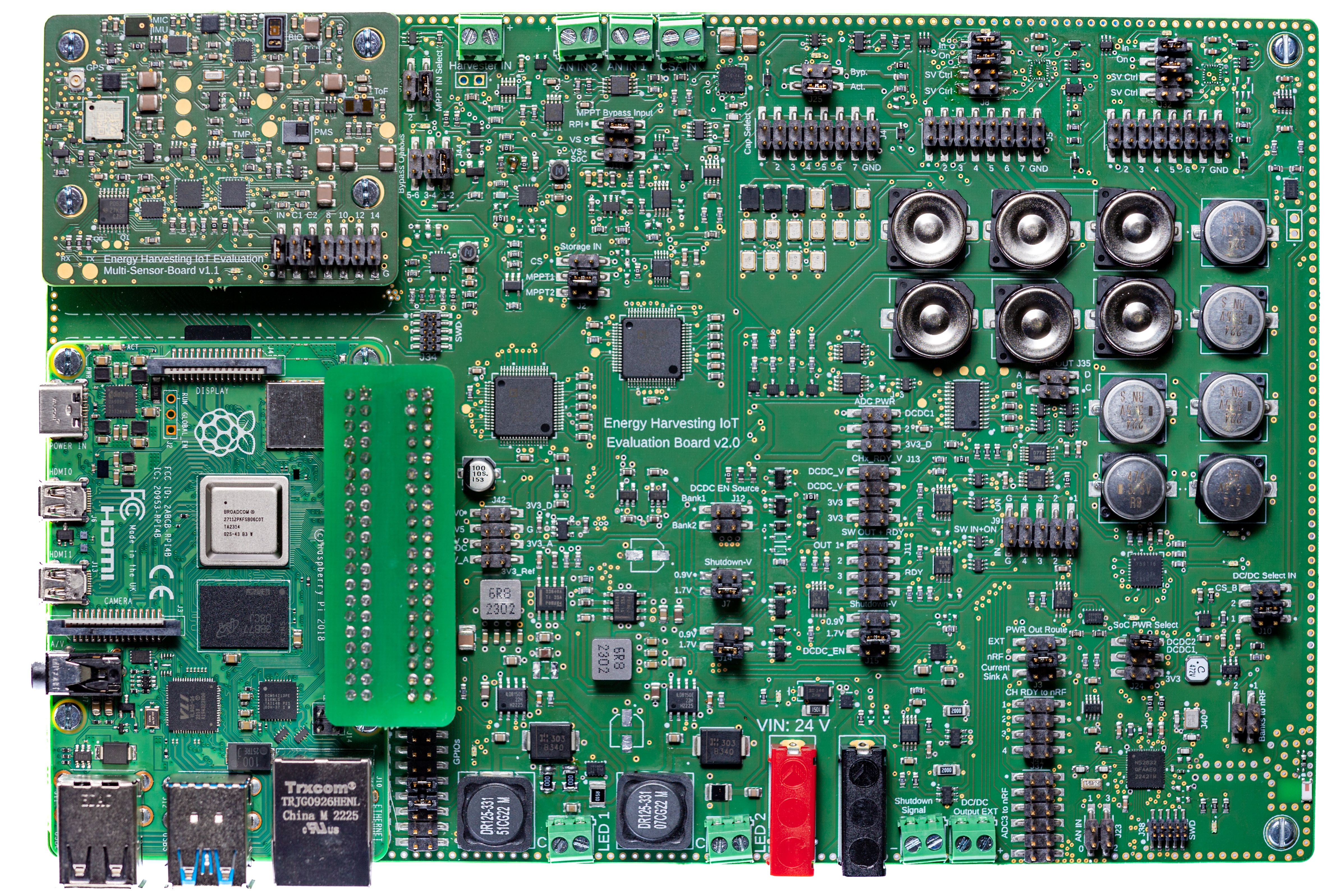}
        \vspace{1pt}
        \caption{The \platformname{} evaluation platform.}
        \label{fig:Platform:Picture}
    \end{subfigure}
    \caption{The \platformname{} evaluation platform. {\it EStacker enables (i) repeatable evaluation of EH-based IoT systems and (ii) accurately attributing energy consumption to the IoT system components and application tasks through energy stacks.}}
\end{figure}

We now present EStacker in detail. Figure~\ref{fig:Platform_Block_Diagram} provides an overview of EStacker's high-level system architecture, while Figure~\ref{fig:Platform:Picture} shows a picture of the actual EStacker evaluation platform. The design is centered around a state-of-the-art {\it EH IoT System}, as shown in Figure~\ref{fig:EH_IoT_Hardware_Arch} (marked in red). The components of the {\it Introspection and Control Unit (ICU)} are shown in green and are responsible for capturing measurement data and emulating the energy environment and events. EStacker also supports more idealized setups in which (i) the harvester and MPPT are emulated or (ii) the SSS is emulated, and the components required to support these modes are shown in orange in Figure~\ref{fig:Platform_Block_Diagram}.

\subsection{The EH IoT System in EStacker}

We designed EStacker's EH IoT System to offer the flexibility of selecting different design points with the same underlying introspection subsystem. Therefore, the ESS consists of two different MPPTs, a flexibly dimensionable energy storage, and two different output DC/DC converters. If desired, these can alternatively be replaced by custom shields, following a modular approach (see, e.g., \cite{bakar_protean_2023, geissdoerfer:2019:shepherd, hester:2017:flicker}). The SSS consists of an ultra-low-power SoC with Bluetooth low energy (BLE) support and a custom sensor board. As an alternative to using the actual SSS, \platformname{} also supports the emulation of arbitrary consumption profiles, similar to \enquote{dummy} or \enquote{smart} loads supported by Shepherd~\cite{geissdoerfer:2019:shepherd} and EHTestbed~\cite{sigrist_environment_2021}. On the input side, an energy harvester with direct-current output (e.g., a solar cell or a thermoelectric generator) can be connected externally to the MPPT. 
We will explain the specific platform configuration used for our experiments in greater detail in Section~\ref{Sec:Exp_Setup}.

\subsection{The Introspection and Control Unit (ICU)}

\platformname{} supports two different state-of-the-art approaches for emulating realistic energy inputs. 
The first method is inspired by  EHTestbed~\cite{sigrist_environment_2021} and controls the energy environment of a real energy harvester. The solar energy emulation option that we focus on in this work relies on irradiance data recorded at a certain location, for example retrieved from comprehensive global meteorological datasets such as JRC-SARAH3 and JRC-ERA5 via PVGIS~\cite{EU:2025:PVGIS}. The {\it Data Aquisition and Control Unit (DACU)}, which is the main unit within the ICU, then uses the recorded irradiance to linearly control a high-power LED from which the solar panel of the EH IoT System then harvests energy.
Alternatively, the energy harvester itself can be emulated. This approach is based on recording and replaying a time series of IV-curves, also referred to as IV-surfaces~\cite{hester:2014:ekho, furlong_realistic_2016, hester:2017:realistic}. They capture both the variation in the energy environment and the electrical characteristics of the energy harvester used to record the data. 
\platformname{} supports replaying IV-surfaces by using an ADC to measure the output voltage seen by the emulated harvester and a programmable current source to output an emulated harvester current based on the recorded IV-surface and the measured voltage. In this way, EStacker can easily support energy sources other than solar energy.

Many IoT applications change their behavior depending on external events detected by their sensor(s). Examples include gesture-activated remote controls, temperature monitors with alarms, or smart parking applications~\cite{colin:2018:reconfigurable, al-turjman:2019:smart}. Evaluating reactive applications in a fair and repeatable manner requires exposing all the configurations that the developer wants to compare to the same events at the same time. EStacker supports this through event traces. During evaluation, the DACU parses the (optional) event trace alongside the ambient energy trace and signals the SoC via a general-purpose I/O pin whenever an event is pending. The system under test thus continues to use its actual sensor(s) for sampling, but instead of taking action based on the sampled data, it checks the event signal immediately after sampling. Combining a real sensor with an emulated event trace is hence an attractive compromise as it yields realistic sensor energy consumption while at the same time ensuring that all explored configurations are exposed to the same events at the same time. Moreover, it eliminates the need to emulate the physical environment of each sensor type (e.g., motion, temperature, or pressure), thereby significantly reducing the complexity of the evaluation setup.

To generate energy stacks, the ICU monitors the power consumption of the key units of the EH IoT System through the {\it Data Collection and Pre-Processing Unit (DCPU)}.
The voltages are either scaled first or measured directly with high-impedance input ADCs, while the currents are measured via shunt resistors and current sense amplifiers.
We find that the EH IoT System can exhibit short peaks of (relatively) high power consumption, and EStacker must therefore sample the ADCs (very) frequently; this observation is in line with prior work~\cite{kazdaridis:2020:eprofiler, sigrist_environment_2021}. The DACU cannot achieve sufficiently high sampling frequencies, and we hence added the DCPU to push the sampling frequency up to 800\,kHz.
The DCPU interfaces with the ADCs to receive the raw sample data, converts them into their corresponding voltage or current values, and then provides their average values to the DACU at a much lower frequency (5\,Hz). An additional benefit of the data aggregation of the DCPU is that it significantly reduces the amount of trace data that must be communicated and stored.
The SoC uses general-purpose I/O pins to report changes in its current activity, e.g., sampling or communicating, to the DCPU, which in turn enables breaking down SoC energy consumption across application activities. 
The ICU has minimal impact on the EH IoT System because it (i) has its own energy supply and (ii) uses low-leakage introspection components.

\section{ST-SP: Reducing Evaluation Time for EH IoT Systems}

\label{Sec:The_Scheme}

Exposing a battery-less IoT system to a sufficient breadth of energy scenarios can drive evaluation times into the range of (many) months, and any optimization that reduces evaluation time is hence attractive. We now explain how our ST-SP approach reduces the evaluation time of IoT applications with (some degree of) periodicity while retaining accuracy.

\subsection{ST-SP Overview}

\begin{figure}[t]
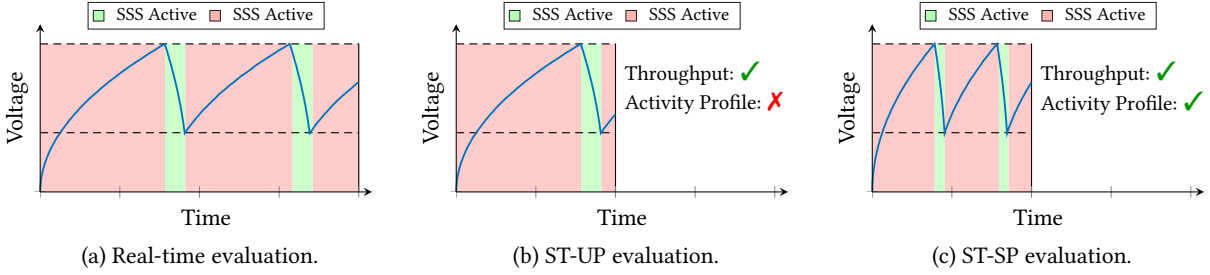

    \begin{subfigure}{0.33\columnwidth}
        \hspace*{0.6cm}\includegraphics{figures/Scaling_Example_RT.tikz}
        \vspace*{-0.1cm}
        \caption{Real-time evaluation.}
        \label{fig:Scheme:Scaling_Example_RT}
    \end{subfigure}
    \begin{subfigure}{0.33\columnwidth}
        \hspace*{0.6cm}\includegraphics{figures/Scaling_Example_STUP.tikz}
        \vspace*{-0.1cm}
        \caption{ST-UP evaluation.}
        \label{fig:Scheme:Scaling_Example_STUP}
    \end{subfigure}
    \begin{subfigure}{0.33\columnwidth}
        \hspace*{0.6cm}\includegraphics{figures/Scaling_Example_STSIOP.tikz}
        \vspace*{-0.1cm}
        \caption{ST-SP evaluation.}
        \label{fig:Scheme:Scaling_Example_STSIOP}
    \end{subfigure}

    \caption{Real-time evaluation compared to 2$\times$ accelerated evaluation with ST-UP and ST-SP.
    {\it Capturing both performance and temporal behavior of real-time execution requires proportionally scaling time, input, and output power.}}
    \label{fig:Scheme:Scaling_Examples}
    \vspace*{-0.3cm}
\end{figure}

The temporal behavior of an EH-based IoT system highly depends on its energy storage size~\cite{colin:2018:reconfigurable}, which is determined, e.g., by its capacitance. Under the same conditions, a larger storage size takes longer time to charge to a voltage that is sufficient for the SSS to start operating but --- once it has reached this threshold --- it can operate for longer, before a minimum voltage threshold is reached and the SSS has to shut down. Typically, these devices spend (much) more time charging than operating, since input power is usually (much) lower than the average power consumption~\cite{colin:2018:reconfigurable, ghasemi_ess:_2023}. Figure~\ref{fig:Scheme:Scaling_Example_RT} shows an example of the energy storage voltage over time for such a device, which was modeled with constant input power for illustration purposes. Red phases indicate charging (SSS off) and green phases mark active periods (SSS operating). The turn-on and turn-off voltage thresholds are displayed as dashed lines.

Figure~\ref{fig:Scheme:Scaling_Example_STUP} shows the result of speeding up the evaluation by a factor of 2 by scaling time but not scaling power, i.e., the scaled time, unscaled power (ST-UP) approach used to reduce evaluation time in EHTestbed~\cite{sigrist_environment_2021}.
Since energy is the product of power and time and throughput is proportional to energy, the throughput under ST-UP is expected to be $\frac{1}{S}$ of the throughput under real-time evaluation when the speed-up factor is $S$, i.e., $\frac{1}{2}$ for the 2$\times$ speed-up in this example.
ST-UP can hence provide reasonable throughput estimates, but it comes at the cost of losing important temporal information. 
Figure~\ref{fig:Scheme:Scaling_Example_STUP} shows that due to the unchanged input and output power, charging and execution times are the same as during real-time execution and re-scaling the accelerated experiment back to the original time axis (by stretching it by the speed-up factor of 2) therefore results in a very different activity profile. More specifically, ST-UP predicts a single active period that is twice as long as each of the two active periods under real-time evaluation.

A better approach is to scale both time and power, i.e., adopt an ST-SP strategy, and thereby retain the real-time balance of supplied and consumed energy when accelerating evaluation. 
In contrast to ST-UP, ST-SP scales input and output power in addition to evaluation time. Figure~\ref{fig:Scheme:Scaling_Example_STSIOP} shows that ST-SP in this way retains the real-time activity profile by exhibiting two active periods that are half as long as under real-time evaluation. Scaling up the ST-SP results by the activity factor thus retains both throughput and activity profile, while ST-UP will only retain throughput.

\subsection{ST-SP: The Details}
\label{Sec:Scheme_Theory}

In the following, we will describe how the ST-SP approach can be applied in practice to speed up the evaluation of battery-less IoT systems. ST-SP strives to keep the amount of supplied (and consumed) energy equal to the amount during real-time execution $E_\text{Real-Time}$, while reducing evaluation time. This can be achieved by proportionally scaling time $t_\text{RT}$ and (average) power $\overline{P}_\text{RT}$ by the same factor $S_\text{TP}$:

\begin{equation}
    E_\text{Real-Time} = \overline{P}_\text{RT} \times t_\text{RT} = \left ( \overline{P}_\text{RT} \times S_\text{TP} \right ) \times \left ( t_\text{RT} / S_\text{TP} \right )
    \label{Eq:Energy_Realtime_Spedup}
\end{equation}

Scaling input power in trace-based evaluation, e.g., with illuminance traces~\cite{sigrist_environment_2021}, power traces~\cite{ghasemi_ess:_2023}, or IV-traces~\cite{hester:2017:realistic, geissdoerfer:2019:shepherd}, is straightforward, as the traces can be directly multiplied by the scaling factor. In the case of solar emulation, this would increase the current flowing through the LED(s), which increases the available energy in the environment.
Increasing power on the consumption side can however be challenging as evaluation involves executing a target IoT application on specific hardware. Fortunately, many IoT applications exhibit (some form of) periodicity, and, in such applications, we can leverage that power consumption changes in response to sampling frequency.

\begin{figure}[t]
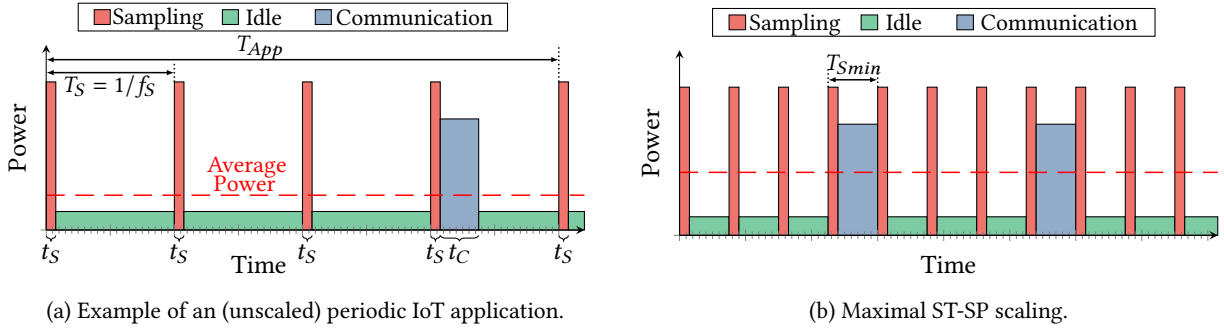

    \begin{subfigure}{0.49\columnwidth}
    \hspace*{0.6cm}\includegraphics{figures/Application_Time_Plot_orig.tikz}
            \caption{Example of an (unscaled) periodic IoT application.}
            \label{Fig:Application_Time_Plot_orig}
    \end{subfigure}
    \hfill
    \begin{subfigure}{0.49\columnwidth}
        \vspace*{0.3cm}\hspace*{0.6cm}\includegraphics{figures/Application_Time_Plot_spedup.tikz}
        \caption{Maximal ST-SP scaling.}
        \label{fig:Application_Time_Plot_orig_spedup}
    \end{subfigure}
    \caption{Example of task execution in a periodic IoT application. {\it ST-SP relies on increasing the average power consumption of the application by increasing its sampling frequency.}}
    \label{fig:Application_Time_Plot}
\end{figure}

Figure~\ref{Fig:Application_Time_Plot_orig} illustrates a typical periodic IoT application, which takes samples at fixed time intervals $T_\text{S}$, which corresponds to the reciprocal of the sampling frequency $f_\text{S}$. After it has collected a certain number of samples $n_\text{S} \geq 1$, it communicates them to a back-end system. For subsequent calculations, we will refer to $t_\text{A}=n_\text{S} \times t_\text{S}+t_\text{C}$ as the total active time resulting from the sum of all sampling times $t_\text{S}$ and the communication time $t_\text{C}$ within the application period $T_\text{App}$. We define the average power consumption of active tasks $\overline{P}_\text{A}$ via their corresponding energy consumption in relation to the total execution time. Whenever the application is not actively executing one of the previously mentioned tasks, the SoC is idle and usually enters a sleep mode to conserve energy. Analogous to $\overline{P}_\text{A}$, we denote $\overline{P}_\text{I}$ as the average power consumption during idle or sleep periods (i.e., the total idle energy divided by total execution time).
The average power consumption of the entire application is therefore $\overline{P}_\text{App} = \overline{P}_\text{A} + \overline{P}_\text{I}$.

As shown in Equation~\ref{Eq:Energy_Realtime_Spedup}, the goal is to increase the application's average power consumption by the scaling factor $S_\text{TP}$ such that:

\begin{equation}
\overline{P}_\text{target} = S_\text{TP}\times \overline{P}_\text{App} = S_\text{TP}\times\left(\overline{P}_\text{A} + \overline{P}_\text{I} \right).
\label{eq:P_target}
\end{equation}
Since the average power consumption during sampling and communication is usually (much) higher than while idle, we can simply scale the average power consumption by increasing the sampling frequency $f_\text{S}$. A key benefit of changing only this single parameter is that the program structure itself can be left unmodified.

While increasing $f_\text{S}$ usually leads to an increase in the energy consumption of the SSS, they are not proportional. The reason for that is that the SoC spends less time idling as a result of active tasks being executed more frequently, and therefore the average idle power $\overline{P}_\text{I}$ decreases. More specifically, in each application period, the application is idle whenever it is not executing sampling or communication tasks (with their total time $t_\text{A}$). For an unscaled application, the idle time is $T_\text{App}-t_\text{A}$. If the sampling frequency is scaled up by $S_\text{f}$, this leads to a reduction of idle time to $T_\text{App}-t_\text{A}\times S_\text{f}$. Consequently, applying the scaling factor $S_\text{f}$ to a periodic application leads to (i) a proportional increase in average active power and (ii) a reduction in average idle power by the idle time ratio before and after scaling. Hence, the resulting average power consumption of the SSS is:
\begin{equation}
\overline{P}_\text{scaled} = S_\text{f} \times \overline{P}_\text{A} + \frac{T_\text{App}-t_\text{A} \times S_\text{f}}{T_\text{App}-t_\text{A}} \times \overline{P}_\text{I}.
\label{eq:P_scaled}
\end{equation}
Equation~\ref{eq:P_scaled} shows that setting $S_\text{f}$ to the time and power scaling factor $S_\text{TP}$ does not lead to the desired power scaling (see $\overline{P}_\text{target}$ in Equation~\ref{eq:P_target}). To calculate the correct scaling factor for the average power consumption of an IoT application using $S_\text{f}$, given a target speed-up factor $S_\text{TP}$, we can equate Equation~\ref{eq:P_target} and \ref{eq:P_scaled} and solve for $S_\text{f}$:
\begin{equation}
S_\text{f} = \frac{S_\text{TP} \times \left( \overline{P}_\text{A} + \overline{P}_\text{I} \right) \left( T_\text{App}-t_\text{A} \right) - T_\text{App} \times \overline{P}_\text{I}}{\left( T_\text{App}-t_\text{A} \right) \times \overline{P}_\text{A} - t_\text{A} \times \overline{P}_\text{I}}.
\label{eq:Sf}
\end{equation}
ST-SP can increase the average power consumption up to the point at which active tasks are executed back-to-back as illustrated in Figure~\ref{fig:Application_Time_Plot_orig_spedup}. In this case, the minimum sampling period is $T_\text{Smin} = t_\text{S} + t_\text{C}$.

\subsection{Using ST-SP with EStacker}
\label{SubSec:Scheme_Steps}

\begin{figure}[t]
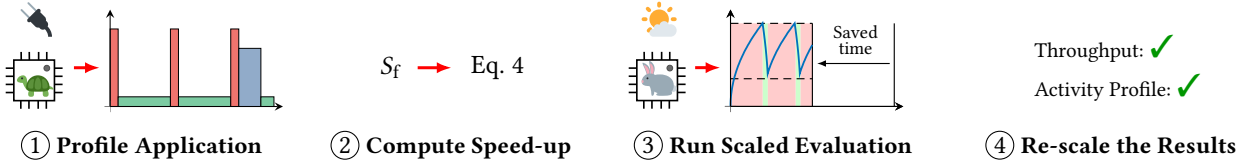

\begin{subfigure}{0.47\columnwidth}  
    \hspace*{1.45cm}\includegraphics{figures/STSP_Implementation_Steps_horizontal_1.tikz}
\end{subfigure}
\begin{subfigure}{0.47\columnwidth}  
    \hspace*{1.85cm}\includegraphics{figures/STSP_Implementation_Steps_horizontal_2.tikz}
\end{subfigure}
\caption{The 4-step process for evaluating a given IoT application with ST-SP.}
\label{fig:Scheme:Steps}
\end{figure}

Figure~\ref{fig:Scheme:Steps} illustrates the 4-step process required to apply ST-SP to a given periodic IoT application.
The first step is to profile the application to capture its power consumption profile and extract the parameters required to compute Equation~\ref{eq:Sf}, see \circled{1}{0.8} in Figure~\ref{fig:Scheme:Steps}. For this purpose, we deploy the application on \platformname{} and switch its power supply from the ESS to a constant 3.3\,V supply. Since ST-SP focuses on periodic applications, the operating power consumption of the SSS with an external power supply yields an upper bound on the performance that can be achieved when powered by energy harvesting.
While the application is running, we record voltage and current data and log the application's activities over several application periods. In addition, we track the throughput of the application (i.e., the number of transmitted bytes), which we will use later in step~\circled{4}{0.8} to calculate the predicted throughput from the scaled experiment.

The second step is to compute the speed-up factor $S_\text{f}$. By analyzing the recorded data, the average power and timing information for solving Equation~\ref{eq:Sf} can be found (see~\circled{2}{0.8}). In order to determine the application's maximum speed-up factor $S_\text{f}$, the time and power scaling factor $S_\text{TP}$ can be increased until the scaled sampling time is equal to the longest active period, which in many cases consists of the sum of the sampling time and communication time. Therefore, the following must apply: $S_\text{f} \leq T_\text{S} / \left (t_\text{S} + t_\text{C} \right )$.

Third, the scaled experiment is executed on EStacker.
Once the application speed-up factor $S_\text{f}$ is calculated for a time and power scaling factor $S_\text{TP}$, the experiment can be configured as follows to speed up the evaluation while retaining accuracy (step~\circled{3}{0.8}). On the input side, the energy traces (e.g., irradiance traces for solar evaluation) are replayed at $S_\text{TP}$ times the real-time speed. To retain the energy of the original environment, its intensity (e.g., the LED current for solar emulation) has to be scaled up by the same factor. To account for the higher average power on the input side, the application's power consumption also has to be scaled up by increasing its sampling frequency by $S_\text{f}$. If further reduction in evaluation time is required, a possible optimization is to fast-forward the playback of the input trace when its value is zero and the SSS is off. This scenario often occurs during the night with solar-powered systems. The effect of this optimization on accuracy depends mainly on the leakage of the ESS components. We evaluate this \textit{skip-nights} optimization in Section~\ref{Sec:STSP_Results} and show that it yields similar accuracy to pure ST-SP.

Once the experiment has completed, we predict the activity profile of the application under real-time evaluation by mapping it back to the original time axis (step~\circled{4}{0.8}). Specifically, this can be achieved by multiplying each time step of the scaled experiment by $S_\text{TP}$. 
The measured throughput, however, differs from what could have been expected from the original application running in real-time because the application was modified for the accelerated experiment. In order to derive the throughput of the original application, we can use the following observation. During the application profiling (step~\circled{1}{0.8}), the unscaled application has produced a certain throughput $\theta_\text{Profiling}$ during its runtime $t_\text{Profiling}$. Due to the fact that in periodic applications, throughput is proportional to runtime, the real-time throughput $\theta_\text{RT}$ can now be calculated based on the total application runtime (active time) $t_\text{Exp}$ during the accelerated experiment by:
\begin{equation}
\theta_\text{RT} = \frac{S_\text{TP} \times t_\text{Exp}}{t_\text{Profiling}} \times \theta_\text{Profiling}.    
\label{Eq:Rescale_Performance}
\end{equation}

\subsection{ST-SP: Overheads and Limitations}

The overheads of applying ST-SP during evaluation lie in (i) the need to profile the original application, (ii) calculating the appropriate sampling frequency scaling factor $S_f$, and (iii) modifying the original application so that its average power consumption under ST-SP is increased. In the profiling step, the power consumption of the SSS must be monitored over at least the longest period occurring in the application, which in our benchmarks is maximally 120 seconds. We decided to profile each application for 1 hour in order to obtain more accurate figures by averaging them. Considering that the total evaluation time for slow-changing energy environments such as outdoor solar can easily add up to several days and weeks, the profiling overhead is hence negligible. 
Determining the sampling frequency scaling factor $S_f$ involves solving Equation~\ref{eq:Sf}. When running on EStacker, this is straightforward as it provides all the required power and timing measurements. The final step before running the actual experiment is to modify the original application to run under ST-SP. Given that the application's source code is available, this simply amounts to changing the task execution frequency according to $S_f$, which in many cases only requires changing a single parameter. 

ST-SP relies on increasing the application's average power consumption under accelerated evaluation, and this is typically achievable if the application exhibits some form of periodicity. ST-SP handles applications that react to events by speeding up the replay of the event traces in sync with the input energy trace to maintain correct temporal alignment. ST-SP cannot be applied to purely event-driven applications, i.e., applications that only execute tasks when external events occur, because this would require speeding up the event trace more than the energy trace and break temporal alignment. Such applications can however still be evaluated on \platformname{} in real time. 
ST-SP's maximum achievable speed-up is limited by (i) the longest active period in the target application, and (ii) the highest power density that can be produced in the emulated energy environment, e.g., a maximum power density of $220\,\text{W/m}^2$ in EStacker's current solar energy setup.

\section{Experimental Setup}

\label{Sec:Exp_Setup}

\renewcommand{\arraystretch}{1.2}   

\begin{table}[]
\centering
\caption{Benchmarks.}
\label{tab:benchmarks}

\begin{tabular}{ll|ccc|c|c|c|c}
\hline
\multirow{2}{*}{Type} & \multicolumn{1}{c|}{\multirow{2}{*}{Abbr.}} & $T_S$ & \multicolumn{1}{l}{$d_S$} & $n_S$ & \multirow{2}{*}{\begin{tabular}[c]{@{}c@{}}SoC Post- \\ Processing\end{tabular}} & $S_\text{I}$ & \multirow{2}{*}{$S_\text{TP}$} & \multirow{2}{*}{$S_\text{f}$} \\
 & \multicolumn{1}{c|}{} & (s) & (B) &  &  & (\%) &  &  \\ \hline
\multirow{2}{*}{Temperature} & TMP1 & 20 & 12 & 1 & - & 2.0 & 3 & 3.4 \\
 & TMP2 & 20 & 12 & 30 & Average & 1.5 & 2 & 2.6 \\
Acceleration & IMU & 60 & 180 & 5 & FFT & 1.5 & 7 & 10.9 \\
Presence & PMS & 60 & 12 & 1 & - & 1.5 & 6 & 10.9 \\
Distance & TOF & 120 & 12 & 1 & - & 2.0 & 10 & 12.3 \\
Health & BIO & 120 & 192 & 1 & - & 3.0 & 10 & 10.6 \\ \hline
\end{tabular}
\end{table}

\renewcommand{\arraystretch}{1}  
\setlength{\tabcolsep}{5pt}

\subsection{Benchmarks}
\label{subsec:benchmarks}

To showcase the features of EStacker, we have developed a set of benchmarks that cover a wide range of typical IoT applications. Table~\ref{tab:benchmarks} summarizes the configuration of the individual benchmarks used in this paper, in which the first three parameters capture the key characteristics of the benchmarks. More specifically, $T_S$ describes the sampling period, $d_S$ is the number of bytes transmitted during each communication event, while $n_S$ is the communication incidence~\cite{ghasemi_pes:_2023}, i.e., how many sampling events take place before data is transmitted to the back-end system. All benchmarks use Bluetooth Low Energy to transmit data to our back-end system (i.e., the DACU on \platformname{} as previously shown in Figure~\ref{fig:Platform_Block_Diagram}).
The benchmarks \textsf{TMP2} and \textsf{IMU} apply additional post-processing to the sensor samples, such as an FFT or the calculation of an average sample value. The remaining benchmarks directly transmit the sampled raw data. 
Since the applications (in combination with their respective sensor) consume different amounts of energy, we scaled our irradiance traces by $S_\text{I}$ to supply the system with the required amount of energy. This is equivalent to a change in the size (area) of the solar cell~\cite{ghasemi_pes:_2023}.  

We obtained the maximum achievable time scaling factor $S_\text{TP}$ for the benchmarks by first profiling them individually for 1 hour on \platformname{}, using a fixed 3.3\,V energy supply (see step~\circled{1}{0.8} in Section~\ref{SubSec:Scheme_Steps}). We then computed the sampling frequency scaling factor $S_\text{f}$ using Equation~\ref{eq:Sf} for increasing integer values of $S_\text{TP}$ until $S_\text{f} > T_\text{S} / \left (t_\text{S} + t_\text{C} \right )$, which marks the point at which a task overlap would occur (see Section~\ref{Sec:The_Scheme}). The maximum speed-up of the first four benchmarks was limited by this constraint. For the benchmarks \textsf{TOF} and \textsf{BIO}, however, the maximum speed-up was limited by the maximum output power of \platformname{}'s LED driver.

Our benchmarks rely on real sensors (implemented on the sensor board) that communicate with the SoC via an I²C bus. The benchmarks \textsf{TMP1} and \textsf{TMP2} use a temperature sensor (TMP117~\cite{TMP117}) and represent temperature monitoring applications with high and low reporting frequencies, respectively. The sampling frequencies $1/T_S$ are identical in both cases but \textsf{TMP1} transmits every measurement sample, whereas \textsf{TMP2} only transmits the average temperature value of $n_S$ consecutive samples. 
The application \textsf{IMU} uses an inertial measurement unit (ICM-42670-P~\cite{ICM-42670-P}) to retrieve acceleration data. It represents an example of a predictive maintenance application that collects 256 acceleration samples at 400\,Hz every $T_S$ seconds, calculates an FFT, and stores the 3 highest amplitudes and their corresponding frequencies. After $n_S$ FFTs have been calculated, the data consisting of the maximum amplitudes and frequencies of each FFT are transmitted.
The benchmark \textsf{PMS} is based on an infrared presence and motion sensor (STHS34PF80~\cite{STHS34PF80}). This application is typical for a home automation scenario where for example room heating or cooling is controlled based on the presence of people. 
The application \textsf{TOF} uses a time-of-flight sensor (VL53L1X~\cite{VL53L1X}) for distance measurements as used for example for liquid level monitoring. This benchmark activates the sensor every $T_S$ seconds, measures the distance and transmits the results immediately to the back-end system.
The benchmark \textsf{BIO} uses a biometric sensor hub (MAX32664~\cite{MAX32664}) in combination with a pulse oximeter and heart-rate sensor (MAX30101~\cite{MAX30101}). The application is configured with a low sampling frequency, suitable for low-power sleep monitoring. 

All benchmarks employ a state-of-the-art just-in-time checkpointing scheme~\cite{maeng:2019:supporting}, in which a low-power voltage supervisor (XC6135C17C~\cite{XC6135}) is used to monitor the voltage of the energy storage, and the SoC only stores new application state to non-volatile memory if the voltage is too low. The threshold voltage was set at 1.7\,V to ensure sufficient margin for short voltage drops caused by the equivalent series resistance of the capacitors~\cite{ruppel:2022:architectural}.

\begin{figure}[t]
    \begin{center}
        \hspace{0.9cm}\includegraphics{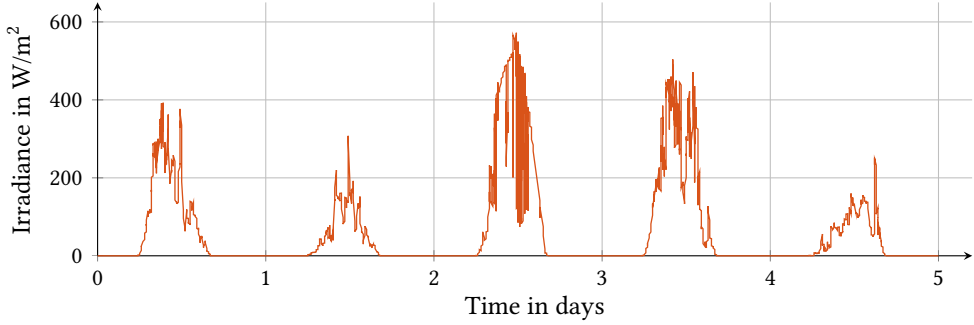}\vspace*{-0.2cm}
    \caption{The selected outdoor solar trace, recorded from 20.10.2022 - 24.10.2022 in southern Germany (station 1)~\cite{dittmann:2024:pvlive}. 
    {\it The trace reflects a challenging energy environment, with high variation both within days and between days.}
    }
    \label{fig:autumn_solar_trace}
    \end{center}
\end{figure}

\subsection{EStacker Configuration}

We configure \platformname{} with a SANYO AM-5412 solar cell~\cite{am-5412} that is connected to a TI BQ25504~\cite{BQ25504} boost converter with integrated MPPT. This particular MPPT is highly inefficient in its cold-start mode which it enters whenever the energy storage voltage is below 1.77\,V. We therefore implement a dynamic bypass similar to prior work~\cite{colin:2018:reconfigurable} which connects the energy storage directly to the solar cell until it has charged to 1.8\,V. When the voltage exceeds this threshold, the bypass is deactivated and the MPPT is enabled until the energy storage voltage drops below 1.6\,V. The integrated boost converter charges the energy storage to a maximum voltage of 2.9\,V.
The energy storage itself consists of several supercapacitors connected in parallel, which have a total capacitance of 2.2\,F. Before each experiment, \platformname{} (dis-)charges the energy storage to 0.75\,V to ensure an identical initial state and thus a fair comparison. Connected to the energy storage, a MAX17223~\cite{MAX17223} DC/DC boost converter provides a stable 3.3\,V supply for the SSS. The converter turns on when the capacitors have been charged to 2.0\,V and stays on until their voltage drops below 0.7\,V. The benchmark applications are executed on a Nordic Semiconductor nRF52832~\cite{nRF52832} SoC, which transmits the data via Bluetooth Low Energy to the DACU.

We use \platformname{} to repeatably emulate a solar energy environment based on an outdoor solar irradiance trace recorded over 5 consecutive days in the fall of 2022 in southern Germany~\cite{dittmann:2024:pvlive} (see Figure~\ref{fig:autumn_solar_trace}). We specifically selected these days because the high variance in irradiance both between days and within days creates a challenging environment for the EH IoT system. The environment emulation on \platformname{} relies on an LED driver (ILD8150~\cite{ILD8150}) that receives a pulse-width modulated signal from the DACU, whose duty cycle is proportional to the irradiance in the solar trace. The LED driver in turn controls a high-power LED mounted above the solar cell in a light-proof box. To ensure that the solar energy environment is accurately reproduced, the LED output was calibrated using a Kipp \& Zonen SP Lite2 pyranometer~\cite{Pyranometer_Lite2}.

\subsection{Metrics}

We use throughput as our overall performance metric, which we define to be the number of bytes transmitted to the back-end system; this definition is in line with prior work (e.g., \cite{ghasemi_pes:_2023}).
We measure real-time throughput by executing the application in our baseline, i.e., without scaling time or power, and measuring the number of bytes transmitted to the back-end system. For the time-scaled configurations, we have to predict the throughput of the real-time baseline. For ST-UP, we do this by multiplying the measured throughput under time-scaling by the time and power scaling factor $S_\text{TP}$ as done in prior work~\cite{sigrist_environment_2021}. For ST-SP, we predict throughput following Equation~\ref{Eq:Rescale_Performance}.

\begin{figure}[t]
    \begin{subfigure}{\columnwidth}
        \hspace*{0.46cm}\includegraphics{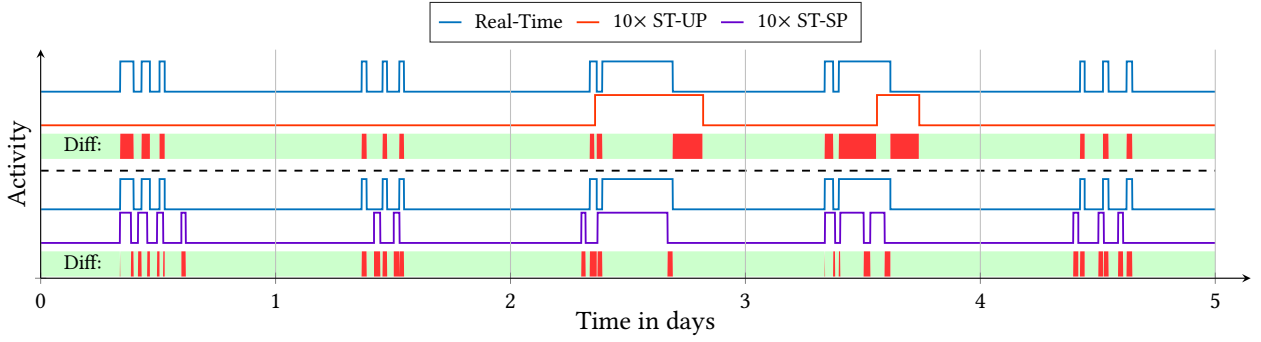}
        \caption{Activity profiles of real-time execution, ST-UP, and ST-SP before applying DTW.}
        \label{fig:Results:BIO_Activity_Original}
    \end{subfigure}
    
    \begin{subfigure}{\columnwidth}
        \vspace*{0.5cm}
        \hspace*{0.46cm}\includegraphics{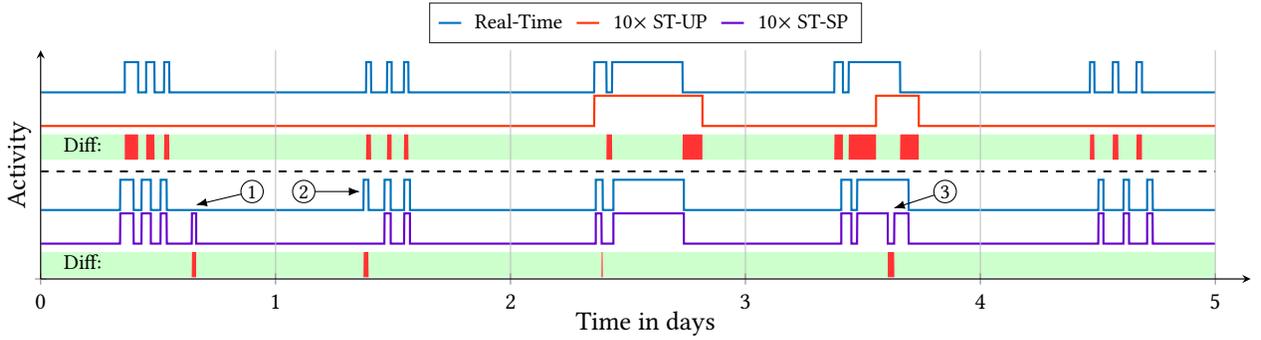}
        \caption{Activity profiles of real-time execution, ST-UP, and ST-SP after applying DTW.}
        \label{fig:Results:BIO_Activity_Warped}
    \end{subfigure}
    \caption{Activity Profile Error (APE) example with the {\sf BIO} benchmark, before and after applying Dynamic Time Warping (DTW) with a window size of one hour. {\it DTW  ensures that APE focuses on significant activity profile deviations.}}
    \label{fig:Results:BIO_Activity_Profile}
\end{figure}

We quantify the {\it Activity Profile Error (APE)} for a scaled evaluation scheme compared to the real-time baseline by dividing execution time into steps (0.2\,s long) and counting the number of time steps in which the activity profiles differ, i.e., the steps in which one profile is on while the other is off and vice versa. 
We then compute APE $\epsilon$ by dividing the resulting difference count $n_d$ by the total number of time steps $n_t$ (i.e., $\epsilon = n_d/n_t$). APE is a number between 0 and 1, where 0 means that activity profiles match exactly and 1 means that they do not match at all.

Figure~\ref{fig:Results:BIO_Activity_Original} shows the activity profiles of the {\sf BIO} benchmark with ST-UP and ST-SP compared to real-time execution. The difference line is green for the time steps where the activity profiles match and red when they do not match, and it illustrates that it is overly restrictive to require an exact activity match. 
First, minor timing differences inflate the APE scores of both ST-UP and ST-SP.
Second, our benchmarks typically run out of energy during the night, which results in prolonged periods where profiles match with both ST-SP and ST-UP, reducing APE scores.
These two effects thus limit the effective resolution of the APE metric and significantly reduce its utility for evaluating time-scaling strategies.
To address this issue, we apply Dynamic Time Warping (DTW)~\cite{sakoe:1978:dynamic} to the activity profiles before computing APE to filter minor activity profile differences;  Figure~\ref{fig:Results:BIO_Activity_Warped} shows the result. APE now captures the fundamental activity profile differences, i.e., ST-SP's additional active period at \circled{1}{0.8}, its missing active period at \circled{2}{0.8}, and its additional shutdown at \circled{3}{0.8}.

\section{Results}
\label{sec:results}

\subsection{Energy Stacks}

\begin{figure}[t]
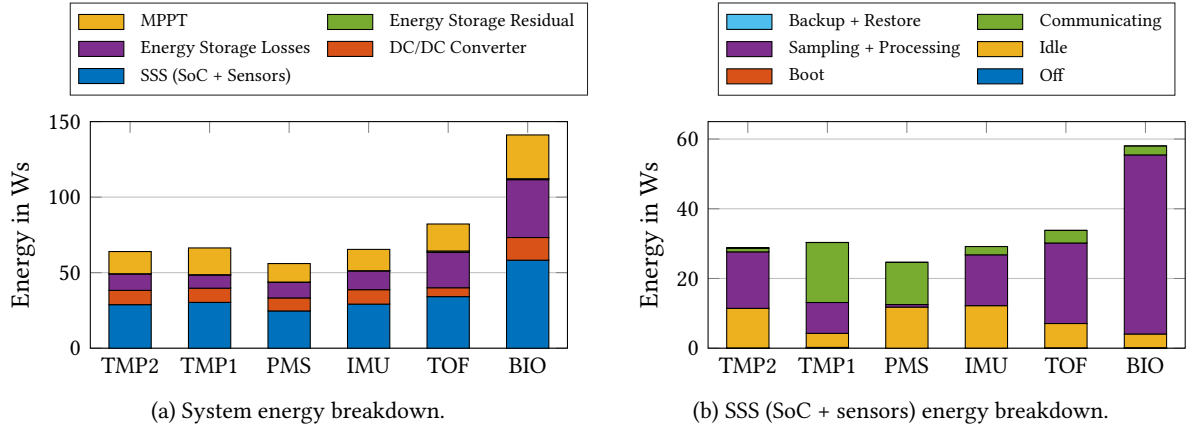

    \begin{subfigure}{0.48\columnwidth}
        \vspace*{0.3cm}\hspace*{1.2cm}\includegraphics{figures/Benchmark_System_Energy_Overview.tikz}
        \caption{System energy breakdown.}
        \label{fig:System_Energy_Overview}
    \end{subfigure}
    \begin{subfigure}{0.48\columnwidth}
    \vspace*{0.3cm}\hspace*{1.4cm}\includegraphics{figures/Benchmark_SoC_Energy_Overview.tikz}
            \caption{SSS (SoC + sensors) energy breakdown.}
            \label{fig:SoC_Energy_Overview}
    \end{subfigure}
    \caption{Energy consumption per IoT system component (Figure~\ref{fig:System_Energy_Overview}) and per SoC task (Figure~\ref{fig:SoC_Energy_Overview}), recorded during real-time evaluation. {\it Our benchmarks cover a range of consumption profiles and thereby span an interesting design space.}}
    \label{fig:Benchmarks_Energy_Overview}
\end{figure}

We now use EStacker to generate energy stacks for our IoT benchmarks in the unscaled baseline configuration in Figure~\ref{fig:System_Energy_Overview}.
The height of the stacks reports the total energy consumed by the IoT system when executing the benchmark in an emulated solar energy environment (see Figure~\ref{fig:autumn_solar_trace}) while the categories explain what the benchmark spends its energy on. Overall, the energy stacks demonstrate that the selected benchmark configurations cover a wide range of total energy consumption and are diverse in terms of how this energy is used.

We now explain the energy categories in detail. 
The \textit{Energy Storage Losses} category is composed of the difference between the input energy into the energy storage and its output energy. Inefficiencies in energy storage are caused, for example, by leakage because the insulation resistance of the capacitors is lower than infinite or losses due to their Equivalent Series Resistance (ESR) being greater than zero. If energy is left in the energy storage at the end of the last day of the solar trace and its voltage has not reached the necessary threshold to turn on the DC/DC converter, \platformname{} discharges the energy storage and attributes the discharged energy to the \textit{Energy Storage Residual} category. If the DC/DC converter is still on when the end of the energy trace is reached, the experiment continues until the energy storage voltage reaches the turn-off voltage threshold and the converter switches off. 
The categories \textit{MPPT} and \textit{DC/DC Converter} capture the energy losses within each of these components. As for the energy storage losses, \platformname{} measures the input and output energy for each component and the losses correspond to their difference. 
The \textit{SSS (SoC\,+\,Sensors)} category describes the amount of energy consumed by the IoT application running on the SoC in combination with the sensor(s).

A key observation from Figure~\ref{fig:System_Energy_Overview} is that despite employing a state-of-the-art IoT system architecture and optimizing the hardware design for low leakage and high efficiency (e.g., by implementing the dynamic MPPT bypass similar to~\cite{colin:2018:reconfigurable}), on average only 44.0\% of the total energy collected by the energy harvester and converted into electrical energy is ultimately usable for the SSS. Instead, on average 20.6\% of the harvested energy was lost in the energy storage, 22.6\% in the MPPT, and 12.8\% in the DC/DC converter. While MPPT and DC/DC converter losses do not vary much across the benchmarks, it can be seen that the relative energy storage losses in the {\sf TOF} and {\sf BIO} benchmarks are significantly higher than for the other benchmarks. The reason for this is ESR-related, as these benchmarks have a comparatively high power consumption during sampling, which leads to approximately 2$\times$ higher ESR losses (28.4\% and 27.1\% for {\sf TOF} and {\sf BIO}, respectively) than with the other benchmarks.

Figure~\ref{fig:SoC_Energy_Overview} zooms in on the {\it SSS (SoC\,+\,Sensors)} category in Figure~\ref{fig:System_Energy_Overview} and breaks down energy consumption across application tasks. Specifically, we use \textit{Off} for time periods in which the DC/DC converter, and by consequence the SSS, is not supplied with power. Because the DC/DC converter's output voltage does not decrease to 0\,V when switched off, the energy in this category is not zero but negligible, i.e., less than 0.4\% of the total energy consumption across our benchmarks. The \textit{Boot} category summarizes the energy consumption that is incurred every time the energy storage voltage reaches the turn-on threshold and the DC/DC converter and SoC boots. Due to the low capacitance on the DC/DC converter's output side, the short time required by the SoC to boot, and the small overall number of (re-)boots (maximum 14 times in the case of {\sf BIO}), this category is also negligible, i.e., less than 0.1\% of the total energy consumption. Across our benchmarks, most energy is spent on \textit{Sampling + Processing} and \textit{Communicating}. The former includes both SoC and sensor energy during sampling (since both are active during this task) and possibly post-processing. 
The SoC's energy consumption while transmitting data to the back-end system via Bluetooth Low Energy is covered by the \textit{Communicating} category. The energy consumed during the process of writing data to non-volatile memory is recorded in the category \textit{Backup + Restore}. Due to the just-in-time checkpointing scheme, which only stores data when a shutdown is imminent, the benchmarks only consume a maximum of 0.4\% of their total energy for this task.

\subsection{Reducing Evaluation Time with ST-SP Results}
\label{sec:results_STSP}

\label{Sec:STSP_Results}

We now turn our attention to demonstrating that ST-SP can speed up the evaluation of periodic IoT applications while retaining accuracy. We consider the following configurations:
\begin{itemize}
    \item {\bf Baseline:} This is the real-time evaluation of each benchmark on EStacker which we use as the golden reference, i.e., the errors of the scaled approaches are relative to this configuration.
    \item {\bf ST-UP:} This is the scaled time, unscaled power strategy from EHTestbed~\cite{sigrist_environment_2021} representing the state-of-the-art.
    \item {\bf ST-SP:} This ST-SP, as described in Section~\ref{Sec:The_Scheme}, which proportionally scales time, input, and output power. 
    \item {\bf ST-SP-SN:} This is ST-SP with the skip-nights optimization.  
\end{itemize}

\begin{figure}[t]
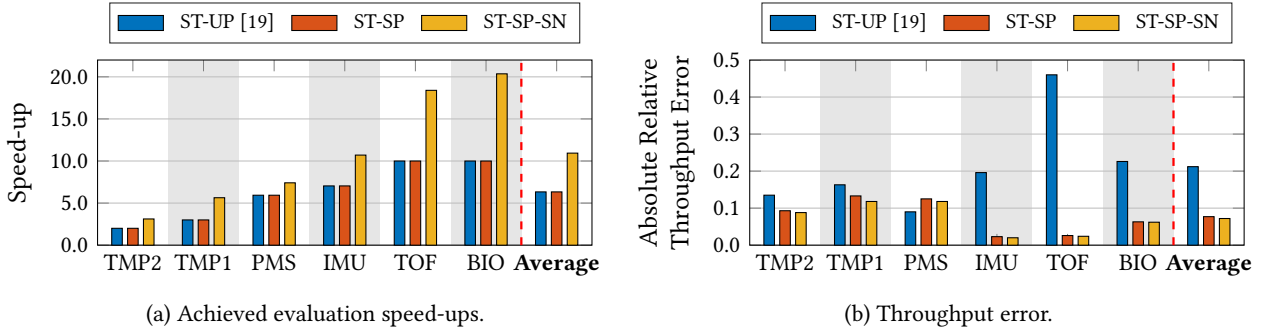

    \begin{subfigure}[b]{0.49\columnwidth}
        \hspace*{1.15cm}\includegraphics{figures/Speedup_Bars.tikz} 
        \caption{Achieved evaluation speed-ups.}
        \label{fig:Results:Speedup_Bars}
    \end{subfigure}
    \hfill
    \begin{subfigure}[b]{0.49\columnwidth}
        \vspace*{0.3cm}\hspace*{1.4cm}\includegraphics{figures/Throughput_Bars.tikz}
        \caption{Throughput error.}
        \label{fig:Results:Throughput_simple}
    \end{subfigure}
    \caption{Evaluation time speed-up and throughput prediction error for ST-UP, ST-SP, and ST-SP-SN, compared to the real-time baseline. {\it ST-SP and ST-SP-SN yield significantly lower average error than ST-UP, i.e., 7.7\% and 7.2\%, respectively, compared to 21.2\%.}} 
    \label{Fig:Res:Throughput_and_Speedup}
\end{figure}

\subsubsection{Evaluation speed-up and throughput error.}

\begin{figure}[t]
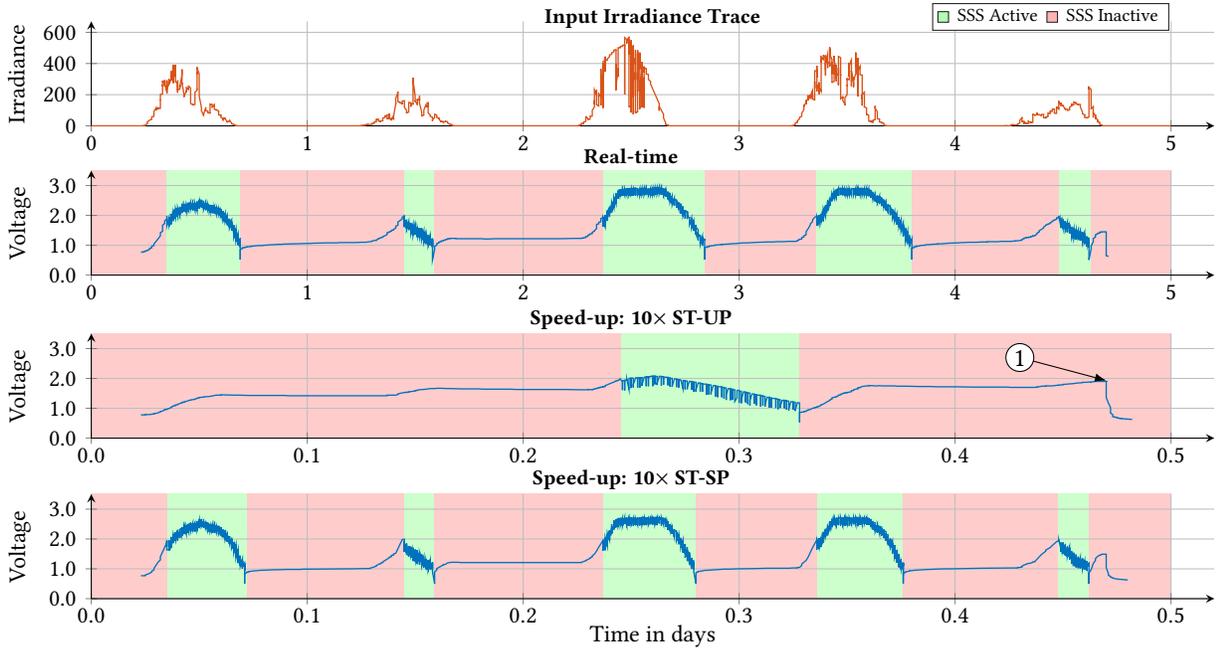

        \hspace*{1.0cm}\includegraphics{figures/TOF2_Time_Plot_RealTime.tikz}\vspace*{-0.15cm}
        \hspace*{1.0cm}\includegraphics{figures/TOF2_Time_Plot_SpedUp.tikz}
        \vspace*{-0.2cm}
        \caption{Irradiance trace versus energy storage voltage for {\sf TOF}. {\it Unscaled input and output power with ST-UP leads to slow charging and discharging, resulting in a large throughput error of 46.0\%. ST-SP's predictions are (much) more accurate (2.6\% error).}}
        \label{fig:Results:TOF2_Time_Plot_RT}
\end{figure}

Figure~\ref{Fig:Res:Throughput_and_Speedup} reports the evaluation speed-up and throughput error for ST-UP, ST-SP, and ST-SP-SN. We run both ST-SP and ST-UP with the same maximum time and power scaling factor $S_\text{TP}$ to ensure a fair comparison (see Table~\ref{tab:benchmarks}) and order the benchmarks by $S_\text{TP}$ from lowest to highest.
ST-UP and ST-SP thus achieve a speed-up equal to $S_\text{TP}$ (on average 6.3$\times$ over the real-time baseline), whereas ST-SP-SN applies the skip nights optimization which yields an average speed-up of 10.9$\times$ compared to the baseline. 

Figure~\ref{fig:Results:Throughput_simple} shows that accelerating the evaluation using the state-of-the-art ST-UP approach leads to an average throughput prediction error of 21.2\%. With the proposed ST-SP and ST-SP-SN approaches, the throughput prediction error was reduced to 7.7\% and 7.2\%, respectively. Figure~\ref{fig:Results:TOF2_Time_Plot_RT} explains the significant throughput error ST-UP incurs on the {\sf TOF} benchmark in Figure~\ref{fig:Results:Throughput_simple}. ST-UP does not scale input power which results in less energy being supplied to the system and ultimately fewer energy storage charge and discharge cycles. For this concrete example, we see that after re-scaling ST-UP to the real-time axis, it approximately takes the first 2.5 days for the energy storage to charge to the required voltage (2.0\,V) to turn on the DC/DC converter. At the end of the solar trace, the energy storage voltage only reached 1.9\,V, which was not sufficient to start a second active SSS period (see \circled{1}{0.8} in Figure~\ref{fig:Results:TOF2_Time_Plot_RT}). 

\begin{figure}[t]
    \vspace{6mm}\hspace*{-0.0cm}\begin{minipage}[b]{0.46\columnwidth}
        \hspace*{0.9cm}\includegraphics{figures/TOF2_Sys_Energy_Stack.tikz}
        \captionof{figure}{{\sf TOF} energy stacks. 
        {\it ST-UP's high residual energy (green) is the specific reason for its 46.0\% throughput error, but the root cause is that it poorly captures {\sf TOF}'s activity profile.}}
        \label{fig:Results:TOF2_Sys_Energy_Stacks}
    \end{minipage}
    \hfill
    \begin{minipage}[b]{0.46\columnwidth}
        \hspace*{1.0cm}\includegraphics{figures/Activity_Profile_Error_Bars_TotTimeNorm.tikz}
        \captionof{figure}{Activity profile prediction error (APE) relative to the real-time baseline. {\it ST-SP and ST-SP-SN on average achieve an APE of 1.4\% and 1.0\%, while ST-UP's average APE is 28.6\%.}}
        \label{fig:Results:Activity}
    \end{minipage}
\end{figure}

The ST-UP energy stack in Figure~\ref{fig:Results:TOF2_Sys_Energy_Stacks} supports this observation by showing a significant amount of unused energy left in the capacitors at the end of the experiment, i.e., the category \textit{Energy Storage Residual} is large for ST-UP and negligible for ST-SP, ST-SP-SN, and the baseline. 
While ST-UP error would likely decrease if we ran our evaluation for a little longer in this specific case, the fundamental issue is that ST-UP is highly sensitive to such effects because it does not retain the application's activity profile. We will discuss this in more detail shortly.  

Figure~\ref{fig:Results:TOF2_Time_Plot_RT} also shows that during real-time execution, the energy storage reaches its maximum attainable voltage of 2.9\,V on days 3 and 4, which is why the MPPT reduces the charging power for a total of 4.7 hours and at times actively sinks current to protect the capacitors from overcharging. This saturation does not occur with ST-UP due to the lower energy input, which is one of the reasons why the MPPT's energy consumption is 23.3\% lower in ST-UP compared to the baseline (see the MPPT category in Figure~\ref{fig:Results:TOF2_Sys_Energy_Stacks}). However, this effect could not compensate for the large amount of unused energy left in the energy storage at the end of the experiment, which led to the high throughput error.

\subsubsection{Activity profile error and sensitivity analysis.}    
Figure~\ref{fig:Results:Activity} reports  Activity Profile Error (APE) for scaled evaluation with ST-UP, ST-SP, and ST-SP-SN. 
ST-UP yields an average APE of 28.6\% which is much higher than the 1.4\% and 1.0\% average APE of ST-SP and ST-SP-SN. The main reason for ST-UP's significantly higher error is again that less energy is effectively supplied to the system (as there is no power scaling), while the energy storage size remains unchanged. 
Although we used a relatively small energy storage size, i.e., none of our benchmarks can operate throughout the night, it takes significantly longer with ST-UP to charge and discharge the capacitors compared to ST-SP. The charging and discharging times with ST-SP are more in line with the real-time baseline because ST-SP proportionally scales evaluation time {\it and} average input and output power. A concrete example can be observed for the {\sf TOF} benchmark in Figure~\ref{fig:Results:TOF2_Time_Plot_RT} where we can see a single long active period with ST-UP while ST-SP's activity profile matches the baseline.

\begin{figure}[t]
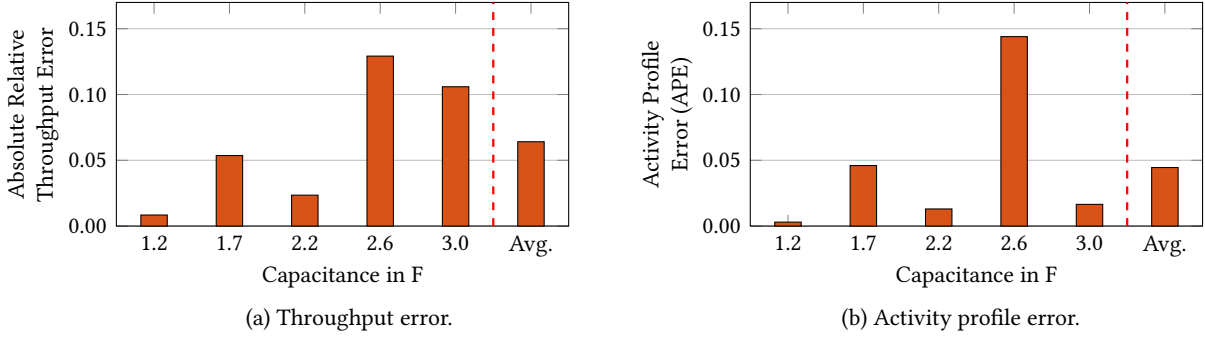

\hspace*{0.7cm}\begin{subfigure}{0.46\columnwidth}
    \hspace*{0.7cm}\includegraphics{figures/IMU2_Capacitor_Variation_Throughput.tikz}\vspace*{-0.1cm}
    \caption{Throughput error.}
    \label{fig:Results:Cap_Variation_Throughput}
\end{subfigure}
\hfill
\begin{subfigure}{0.46\columnwidth}
    \hspace*{1.0cm}\includegraphics{figures/IMU2_Capacitor_Variation_APE.tikz}\vspace*{-0.1cm}
    \caption{Activity profile error.}
    \label{fig:Results:Cap_Variation_APE}
\end{subfigure}
    \vspace*{-0.1cm}
    \caption{ST-SP error for IMU at 7$\times$ speed-up across energy storage configurations. {\it ST-SP retains its accuracy across a range of different energy storage sizes.}}
    \label{fig:Results:Cap_Variation}
\end{figure}

To explore how ST-SP performs across hardware configurations, we reconfigure EStacker's capacitor bank from our default capacitance of 2.2\,F to cover the range from 1.2\,F to 3.0\,F. Figure~\ref{fig:Results:Cap_Variation} reports the throughput error and APE of ST-SP relative to the baseline when running IMU with its maximum 7$\times$ speed-up.
Overall, ST-SP retains its accuracy across the energy storage range with average throughput error and APE of 6.4\% and 4.5\%, respectively.

Errors are however generally higher for the larger energy storage sizes, see the 2.6\,F and 3.0\,F configurations in Figure~\ref{fig:Results:Cap_Variation_Throughput} and the 2.6\,F configuration in Figure~\ref{fig:Results:Cap_Variation_APE}.
The root cause of these errors is that the non-linear interaction between the key ESS components means that the higher input power under ST-SP makes the ESS more efficient, ultimately providing more energy to the SSS. This effect becomes more pronounced when energy storage increases because this increases the ability of the EH system to store and use this additional energy.  

For the 2.6\,F configuration, the ESS provides 13.7\% more energy to the SSS under ST-SP compared to the baseline. This enabled IMU to operate on the second day under ST-SP while remaining off in the baseline, yielding both higher throughput error and higher APE.
The 3.0\,F configuration has a 10.6\% throughput error but APE is only 1.7\%. In this case, the large amount of energy storage enables IMU to remain active for long periods of time once it turns on, and the additional energy available under ST-SP increases the length of all of IMU's active periods, resulting in higher throughput; IMU is on average active for 12.2\% longer under ST-SP than in the baseline. The shape of the activity profile under ST-SP is however similar to the baseline, yielding low APE.

\vspace*{-0.3cm}

\section{Case Studies}
\label{sec:case-studies}

Having performed a general evaluation of EStacker and ST-SP in Section~\ref{sec:results}, we now present two case studies to demonstrate how developers can use them in practice. We first present how we used EStacker's ability to create energy stack profiles to track down and address a challenging performance problem in Section~\ref{subsec:using-estacker} before we leverage the ability of ST-SP to efficiently explore a large design space for a smart parking application in Section~\ref{subsec:exploration-with-st-sp}.

\subsection{Using Energy Stack Profiles for Optimization}
\label{subsec:using-estacker}

\begin{figure}[t]
    \begin{subfigure}{\columnwidth}
        \hspace*{0.1cm}\includegraphics{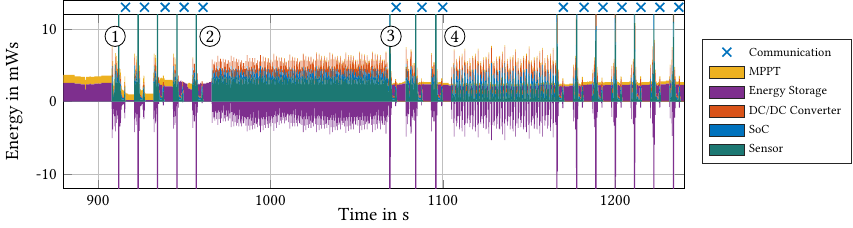}\vspace*{-0.1cm}
        \caption{Energy stack profile on the earlier version of EStacker with the performance problem.}
        \label{fig:CaseStudy:EnergyStacksProblem}
    \end{subfigure}
    \begin{subfigure}{\columnwidth}
        \vspace*{0.4cm}\hspace*{0.1cm}\includegraphics{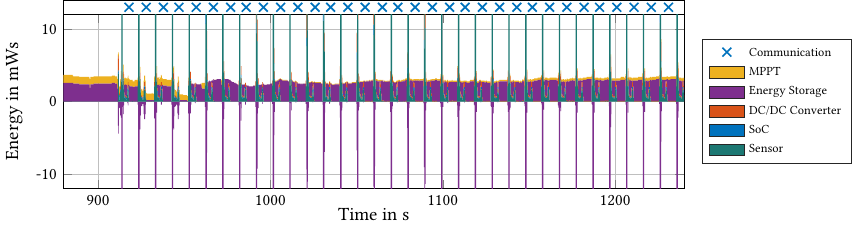}\vspace*{-0.1cm}
        \caption{Energy stack profile when the performance problem has been addressed.}
        \label{fig:CaseStudy:EnergyStacksFix}
    \end{subfigure}
    \caption{Energy stack profile excerpt for {\sf TOF} with 0.2 second time steps demonstrating a performance problem and its solution. {\it Energy stack profiles provide insight into performance issues that are challenging to detect with end-to-end analysis.}}
    \label{fig:CaseStudy:EnergyStacks}
\end{figure}

While we have so far focused on energy stacks that represent the full run of a benchmark, EStacker can also produce energy stack profiles, i.e., a sequence of energy stacks that describe how the energy balance of the system changes over time. Figure~\ref{fig:CaseStudy:EnergyStacksProblem}  shows the energy stack profile for {\sf TOF} on an earlier version of the EStacker platform configured with a total capacitance of $0.54\,\text{F}$. (We showed the energy stack from the full run of this configuration in Figure~\ref{Fig:HighLvl_EvalSetup}.)
Recall that the DACU receives samples at a frequency of $5\,\text{Hz}$ and each energy stacks thus represent 0.2 seconds of application runtime; these stacks are based on averages generated by the DCPU while sampling at $800\,\text{kHz}$. While energy stacks can focus solely on energy consumption, we find that energy stack profiles are more useful when they also show the net change of system energy. As with energy stacks, positive numbers represent consumed energy, and the {\it MPPT}, {\it DC/DC Converter}, {\it SoC}, and {\it Sensor} components are hence always positive. The amount of energy in the {\it Energy Storage} component can however both increase and decrease in a time step. To remain consistent with the other categories, a positive number means that the amount of stored energy increases in this time step. Conversely, a negative number means that the amount of stored energy decreases, see the (large) negative purple spikes in Figure~\ref{fig:CaseStudy:EnergyStacks}.

Figure~\ref{fig:CaseStudy:EnergyStacksProblem} zooms in on the time window from 880 seconds to 1,240 seconds of the execution of the {\sf TOF} benchmark. The first key observation is the periodic and large (green) \textit{Sensor} spikes that occur when the {\sf TOF} benchmark enables the time-of-flight sensor. These spikes can be up to 34.9\,mWs, but we cap Figure~\ref{fig:CaseStudy:EnergyStacksProblem} at 12\,mWs to clearly show the performance problem.
The behavior of the {\sf TOF} benchmark at \circled{1}{0.8} is closer to what the developer would expect with a period of moderate energy consumption while preparing to sample, a spike while the sample is collected, and then a period of moderate energy consumption while the sample is processed and communicated to the back-end system. 
At \circled{2}{0.8} we however see a prolonged green field which means that the sensor continues to consume a significant amount of energy but without successfully sampling and subsequently communicating the sample. The system then temporarily recovers at \circled{3}{0.8} and successfully delivers three samples to the back-end system before entering another unproductive period at \circled{4}{0.8}.

Since the energy stack profiles show that the system does not behave as expected, we analyzed the detailed voltage and current traces provided by EStacker and identified the ESR of the energy storage as the root cause. Although the storage consists of eight supercapacitors connected in parallel, which reduces ESR, the total series resistance is still $R_\text{ESR}=6.9\,\Omega$. The high current drawn by the time-of-flight sensor during sampling (up to $I_\text{SSSmax}=0.15\,\text{A}$) can hence lead to voltage drops of $\Delta V_\text{max} = R_\text{ESR} \times I_\text{SSSmax}=6.9\,\Omega \times 0.15\,\text{A} = 1.04\,\text{V}$ in the worst case when the SSS is supplied solely by the energy storage. 
A voltage drop of this magnitude can easily cause the sensor and SoC to shut down because ESR causes the output voltage to fall below their minimum operating voltage even if there is abundant energy in the capacitors~\cite{ruppel:2022:architectural}. As with many other systems~\cite{denby:2022:tartan, afanasov:2020:batteryless}, our benchmark also uses power gating to minimize the active time of the sensor and thus avoids wasting energy when the sensor is not needed.
At \circled{2}{0.8}, the time-of-flight sensor tries to obtain a sample which due to ESR leads to a voltage drop that causes the SoC to shut down and reboot instead of retrieving the sample from the sensor, turning off the sensor, transmitting the sample, and returning to sleep. The sensor therefore remains powered on, which explains the high {\it Sensor} energy consumption (green fields) in Figure~\ref{fig:CaseStudy:EnergyStacksProblem}. This situation is resolved when the input power of the energy harvester has reached a level at which the SSS receives sufficient energy directly from the harvester for the effect of ESR to become small enough to not cause a shutdown. 

Once the problem was identified, we fixed it by adding low-ESR ceramic capacitors of in total $400\,\mu\text{F}$ to buffer the SSS's supply voltage during short but high current draws from the time-of-flight sensor. Figure~\ref{fig:CaseStudy:EnergyStacksFix} shows the energy stacks over time after fixing the problem.  Notice that the energy stack profile is now much more regular and that the green fields due to the {\it Sensor} category have disappeared. Resolving this issue enables \textsf{TOF} to use more energy on collecting and communicating samples. More specifically, it now communicates 1332 bytes of sample data to the back-end system over 2 days, a $3.3\times$ improvement over the 408 bytes of sample data it was able to deliver before we fixed the issue.

\subsection{Exploring Large Design Spaces with ST-SP}
\label{subsec:exploration-with-st-sp}

In this case study, we target a smart parking scenario~\cite{al-turjman:2019:smart, piccialli:2021:predictive} in which solar-powered sensor nodes are deployed to monitor the occupancy of parking spaces. Each node covers one parking space and sends a message via Bluetooth to the back-end system whenever a car enters or leaves. We derived the application from our \textsf{TOF} benchmark  
which uses a time-of-flight sensor for distance measurements. 
In a real deployment, the application would determine the occupancy status of the parking space based on the measured time of flight, which will be shorter (longer) when the parking space is occupied (free). Recreating this situation in the lab is however impractical as we would need to build a setup in which an object is physically placed at an appropriate distance from the time of flight sensor and then removed at exactly the right times. We therefore use EStacker's event trace feature (see Section~\ref{sec:platform}) to ensure that all configurations see the exact same events.
We randomly generate the event trace based on the distribution we obtained from the popular times graph (Google) of a local shopping center with opening hours from 9:00~AM - 8:00~PM with the highest number of visitors at 14:00~PM. Whenever the SSS is powered, the application takes distance measurements every $T_S = 120\,\text{s}$. 

The system should ideally capture all events, but this requires that it has sufficient energy available when the event occurs. A key design goal for the smart parking application is therefore to select a solar panel and energy storage size to maximize the number of detected parking events. While a larger solar panel harvests more energy, it also requires more physical space and typically increases system cost --- and it is hence important to select an appropriately sized panel. When dimensioning the energy storage, the key trade-off is that while a larger capacitor can store more energy, it also requires being provided with more energy before it reaches a sufficiently high voltage for the SSS to turn on. For this particular application, smaller capacitors do well in the morning --- because the SSS can turn on quickly --- whereas larger capacitors do well in the evening --- because they store enough energy to continue to operate after the sun has set. A further challenge is that the solar panel and energy storage cannot be dimensioned independently because a larger solar panel provides higher power which in turn leads to the capacitor charging faster.

Dimensioning the solar panel and capacitor bank for this application thus requires sweeping a large design space. Since solar energy can vary significantly from day to day, we need to consider multiple days for the results to be trustworthy. We focus on the 5-day outdoor solar trace previously shown in Figure~\ref{fig:autumn_solar_trace}, and consider five solar panel sizes and five energy storage configurations, yielding 25 configurations and an overall sequential evaluation time of 125 days. We have three EStacker boards available which brings the overall evaluation time down to 41.7 days, but this is still more than a month and hence impractical. Applying the ST-SP optimization brought the evaluation time down to roughly one week (7.7 days), yielding an evaluation time speed-up of $5.4\times$.

\begin{figure}[t]
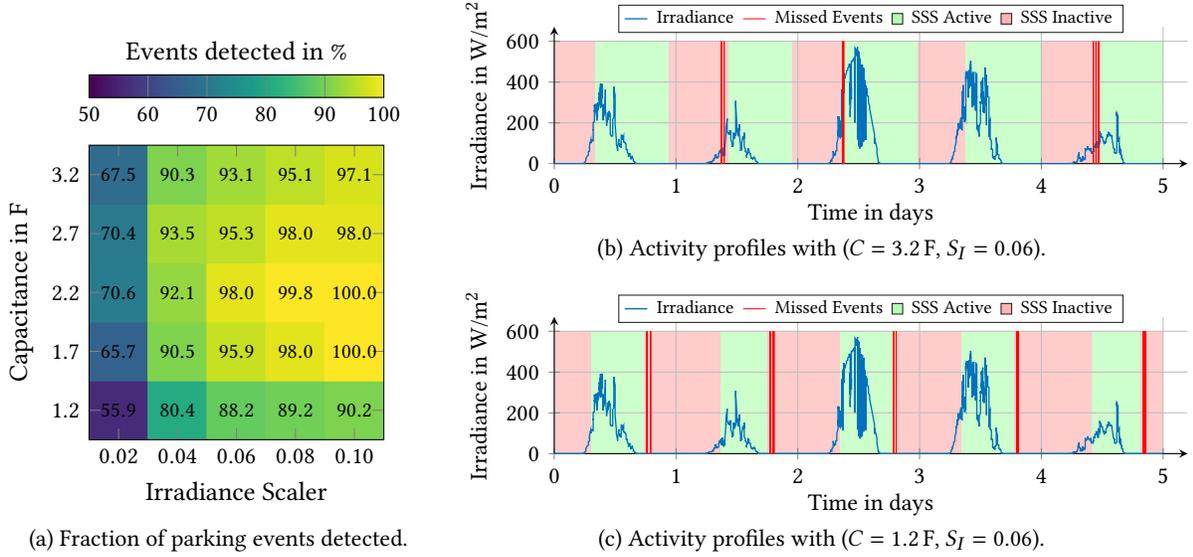


    \begin{subfigure}{0.35\columnwidth}
        \hspace*{-0.0cm}\includegraphics{figures/Parameter_Sweep_Heatmap.tikz}\vspace*{-0.0cm}
        \caption{Fraction of parking events detected.} 
        \label{fig:CaseStudy:Heatmap}
    \end{subfigure}
    \hspace*{0.0cm}
    \begin{minipage}[b]{0.60\columnwidth}
        \begin{subfigure}{\columnwidth}
            \hspace*{1.4cm}\includegraphics{figures/Reactive_App_Time_Clarge.tikz}\vspace*{-0.2cm}
            \caption{Activity profiles with ($C=3.2\,\text{F}$, $S_I=0.06$).}
            \label{fig:CaseStudy:Reactive_Time_Clarge}
        \end{subfigure}
        \begin{subfigure}{\columnwidth}
            \vspace*{0.2cm}\hspace*{1.4cm}\includegraphics{figures/Reactive_App_Time_Csmall.tikz}\vspace*{-0.2cm}
            \caption{Activity profiles with ($C=1.2\,\text{F}$, $S_I=0.06$).}
            \label{fig:CaseStudy:Reactive_Time_Csmall}
        \end{subfigure}
    \end{minipage}
    \caption{Design space exploration with EStacker on a smart parking application. {\it ST-SP makes it feasible to explore larger design spaces because it significantly reduces evaluation time --- in this case from over a month to roughly one week (i.e., from 44.7 days to 7.7 days).}}
    \label{fig:CaseStudy:parking}
\end{figure}

Figure~\ref{fig:CaseStudy:Heatmap} shows the results of the exploration and reports the percentage of parking events detected across our five energy storage configurations $C$ and irradiance scalers $S_I$. Recall from Section~\ref{subsec:benchmarks} that we multiply the real-world solar trace by a constant, which we call the irradiance scaler, to emulate different solar panel configurations. A lower number corresponds to a smaller solar panel and less energy.
Two configurations detect all events: ($C=1.7\,\text{F}$, $S_I=0.1$) and ($C=2.2\,\text{F}$, $S_I=0.1$). These configurations work because they select capacitor sizes that both charge quickly enough to capture the morning events and store sufficient energy to capture evening events with the provided solar panel configuration.  
The reasons why the other configurations miss events are shown in Figure~\ref{fig:CaseStudy:Reactive_Time_Clarge} and~\ref{fig:CaseStudy:Reactive_Time_Csmall}. Figure~\ref{fig:CaseStudy:Reactive_Time_Clarge} focuses on the ($C=3.2\,\text{F}$, $S_I=0.06$) configuration which combines a large capacitor with a moderate solar panel.
The long charge time of the large capacitor leads to missing events in the morning but the large amount of stored energy means that it can operate to the end of the day. 
Figure~\ref{fig:CaseStudy:Reactive_Time_Csmall} focuses on the ($C=1.6\,\text{F}$, $S_I=0.06$) configuration which has a smaller capacitor and the same solar panel, yielding the opposite behavior, i.e., it charges quickly and detects morning events but runs out of energy and misses evening events.

\section{Related Work}

The most related works to ours are EHTestbed~\cite{sigrist_environment_2021}, Ekho~\cite{hester:2014:ekho}, and Shepherd~\cite{geissdoerfer:2019:shepherd} because they, like EStacker, focus on achieving repeatable evaluation of battery-less IoT systems. In contrast to EStacker, they cannot generate energy stacks. This is also the key difference between EStacker and external power monitoring devices such as eProfiler~\cite{kazdaridis:2020:eprofiler} and
RocketLogger~\cite{sigrist_measurement_2017}. EStacker is also inspired by the modular approach taken in platforms for rapid prototyping such as Riotee~\cite{geissdoerfer:2024:riotee}, SuperSensor~\cite{bakar_protean_2023}, REHASH~\cite{bakar_rehash_2021}, Flicker~\cite{hester:2017:flicker}, and FlockLab~\cite{trueb:2020:flocklab, lim_flocklab:_2013}.
The most related work that addressed the challenges in speeding up the evaluation process of battery-less IoT systems is the ST-UP strategy proposed in EHTestbed~\cite{sigrist_environment_2021}. We compared ST-UP quantitatively to our ST-SP and ST-SP-SN optimizations in Section~\ref{sec:results_STSP} and demonstrated that ST-SP and ST-SP-SN yield significantly lower errors.

Looking at battery-less IoT evaluation more broadly, analytical models and simulators are an alternative to the prototype-based evaluation that this work focuses on.
Analytical models such as PES~\cite{ghasemi_pes:_2023} are, unlike hardware prototypes, suitable for exploring of large design spaces and early-stage performance evaluation, as they only require minimal evaluation time. They do, however, rely on average power values and are therefore not suitable for studying scenarios where energy conditions vary significantly. Simulators such as SIREN~\cite{furlong_realistic_2016} or ESS~\cite{ghasemi_ess:_2023} can account for this time-dependent behavior as their evaluation is based on traces that describe the change in the energy environment over time. Although simulators also enable rapid evaluation, they face a fundamental trade-off between model accuracy and evaluation time. Retaining (reasonably) fast evaluation thus typically comes at the cost of abstracting away details that may be important. For example, ESS does not model key electrical properties of the energy supply subsystem, such as ESR~\cite{ruppel:2022:architectural} and operating point-dependent converter losses~\cite{ju:2018:predictive}.

\section{Conclusion}

The performance of battery-less IoT systems greatly depends on the efficiency of its energy supply subsystem and the energy environment at its deployment location.
We hence propose the EStacker evaluation platform to aid developers in evaluating the performance of their battery-less IoT systems in a fair and repeatable manner. The key feature of EStacker is its ability to create energy stacks which break down the energy consumption of the application across hardware units and application activities, thereby explaining what the application spends energy on.
Evaluating battery-less IoT systems can also be time-consuming, especially if the system needs to be exposed to several slow-changing energy scenarios (e.g., outdoor solar environments). 
We hence presented ST-SP which can reduce evaluation time by 6.3$\times$ on average while retaining an average throughput  error of only 7.7\% by 
proportionally scaling evaluation time, input, and output power.
To demonstrate the utility of EStacker and ST-SP, we include two case studies. In the first case study, we use EStacker's energy stack profiles to identify a performance problem in our {\sf TOF} benchmark that, once addressed, improves performance by 3.3$\times$. Second, we use ST-SP to sweep the design space required to identify a favorable combination of solar panel and capacitor size for a smart parking application. ST-SP enables sweeping this design space in roughly one week (7.7 days), while it would have taken over one month (41.7 days) without ST-SP.

\section*{Acknowledgements}
The work presented in this paper has been funded in part by the European Union's Horizon 2020 research and innovation program under the Marie Skłodowska-Curie grant agreement No.~101034240.

\printbibliography

\end{document}